\documentclass{article}
\usepackage{lineno}

\input{tcilatex}
\begin{document}

\title{\hspace{-4cm}{\small \hspace{-4cm}} \\
\relax The relationship between economic growth and environment. Testing the
EKC hypothesis for Latin American countries.}
\author{C. Seri\thanks{{\small University of Roma Tre, Rome, Italy.}} \and %
A. de Juan Fern{\'{a}}ndez\thanks{{\small Corresponding author:
aranzazu.dejuan@uam.es. Universidad Aut{\'{o}}noma de Madrid (UAM): Econom{%
\'{\i}}a Cuantitativa, E-III-307 , Avda. Francisco Tom{\'{a}}s y Valiente 5,
28049 Madrid (Spain).} }}
\maketitle

\begin{abstract}
We employ an ARDL bounds testing approach to cointegration and Unrestricted
Error Correction Models (UECMs) to estimate the relationship between income
and $CO_{2}$ emissions per capita in 21 Latin American Countries (LACs) over
1960-2017. Using time series we estimate six different specifications of the
model to take into account the independent effect on $CO_{2}$ emissions per
capita of different factors considered as drivers of different dynamics of $%
CO_{2}$ emissions along the development path. This approach allows to
address two concerns. First, the estimation of the model controlling for
different variables serves to assess if the EKC hypothesis is supported by
evidence in any of the LACs considered and to evaluate if this evidence is
robust to different model specifications. Second, the inclusion of control
variables accounting for the effect on $CO_{2}$ emissions is directed at
increasing our understanding of $CO_{2}$ emissions drivers in different
countries. The EKC hypothesis effectively describes the long term
income-emissions relationship only in a minority of LACs and, in many cases,
the effect on $CO_{2}$ emissions of different factors depends on the
individual country experience and on the type and quantity of environmental
policies adopted. Overall, these results call for increased environmental
action in the region.



Acknowledgment: The first author acknowledges financial support from
Erasmus+/KA1 Grant number 2019/109282 and the second author acknowledges
financial support from the Spanish Ministry of Science and Technology Grant
ECO2015-70331-C2-1-R. and Spanish Government Project PID2019-108079GBC22.
\end{abstract}

\bigskip

\bigskip

\newpage

\emph{Declarations:}

\begin{enumerate}
\item \emph{Ethics approval and consent to participate }\ Not applicable

\item \emph{Consent for publication} \ Not applicable

\item \emph{Availability of data and materials}: \ The data used in this
manuscript are available at the sources included in table 1 of the
manuscript and are also available from the authors upon request.

\item \emph{Competing interests}: The authors declare that they have no
competing interests.

\item \emph{Funding: }

C. Seri acknowledges financial support from Erasmus+/KA1 Grant number
2019/109282.

A de Juan acknowledges financial support from the Spanish Ministry of
Science and Technology Grant ECO2015-70331-C2-1-R. and Spanish Government
Project PID2019-108079GBC22.

\item \emph{Authors's contribution}.

AdJ has developped the results, doing the estimation of the models

CS has interpreted the results.

Both authors have contributed in writing the manuscript. Both authors read
and approved the final manuscript.
\end{enumerate}

\newpage

\bigskip

\section{Introduction}

The relationship between economic growth and environment has been a matter
of interest for many years and collected academic contributions that date
back to the 1950s. While during the seventies the prevailing view was that
of growth having net adverse environmental impacts (Ehrlich and Holden
(1971); Meadows et al. (1972); Nordhaus (1977)), the stance during the
eighties was more optimistic and was mainly based on the concept of
sustainable development (Brundtland (1987)). The formulation of the
Environmental Kuznets Curve (EKC) hypothesis in the early nineties marked a
significant turning point in this debate and is currently one of its main
focus. The growth-environment relationship ceased being considered a
monotonic one -- whether of positive or negative sign -- and a number of
authors began to argue that the impact of growth on the environment could
change along the course of economic development. According to the EKC
hypothesis, first stated by Grossman and Krueger (1991), in the early stages
of economic growth environmental degradation and pollution increase, but
beyond some level of income per capita the trend reverses with additional
income growth leading to environmental improvement. In analogy with the
relationship between income and income inequality described by Kuznets
(1955), the relationship between economic growth and environmental
degradation could thus be described as an inverted U-shaped curve, hence the
name. A number of elements related to the process of development -- changes
in the economic structure, technological progress, changes in preferences
and increased environmental awareness, among others -- would be at the basis
of such a relationship.

Shafik and Bandyopadhyay (1992) provided a first empirical confirmation of
the hypothesis and popularized the concept. Since then a large stream of
empirical literature flourished using a variety of econometric techniques to
test the hypothesis for different countries, environmental variables and
time periods. Despite a massive empirical literature, the results are highly
heterogeneous and given their sensitivity to the samples and variables
chosen, it could be said that the EKC is not so much an empiric regularity
as it has been believed to be\footnote{%
Moreover, even when the EKC is supported by evidence, in some cases the
estimated turning points for income are so high that environmental
conditions will still deteriorate for a long time before income reaches the
level required to revert the trend (Selden and Song (1994); Shafik (1994);
Holtz-Eakin and Selden (1995); Stern and Common (2001)).} . However, despite
many criticisms, the hypothesis still is among the main approaches to the
study of the relationship between growth and the environment (Stern, 2017).
Moreover, its implications for the design of environmental policies and the
increasingly urgent climate crisis call for a better understanding of these
patterns.

Against this background, in this paper we study the relationship between
income and $CO_{2}$ emissions per capita in twenty-one Latin American
Countries (hereinafter LACs) over 1960-2017. We test the EKC hypothesis
employing Autoregressive Distributed Lag (hereafter, ARDL) bounds testing
approach to cointegration based on Unrestricted Error Correction Model
(hereinafter, UECM). We estimate this model separately in each Latin
American country in our sample and controlling for the effect of different
explanatory variables, following a time series approach. Indeed, while many
studies already tested the EKC hypothesis, studies testing the hypothesis in
Latin America are fewer in number (for example, Mart\'{\i}nez-Zarzoso and
Bengochea-Morancho (2003); Poudel et al. (2009); S\'{a}nchez and Caballero,
2019); Zilio and Caraballo (2014)) and the vast majority of them use panel
data approach, despite the superiority of time series techniques to
investigate the existence of the EKC has been claimed for a long time (De
Bruyn et al. (1998); Lindmark (2002); Stern et al. (1996); Unruh and Moomaw,
(1998); Vincent, (1997)). We estimate six different models for each country,
in order to control for different variables. These variables are chosen to
account for the effect of a number of elements discussed in the theoretical
literature as possible causes to increases or reductions of $CO_{2}$
emissions: output structure, commodity dependence, population density,
external relationships (trade and FDI), agricultural land, rural population
and the energy mix. This approach allows us to address two important issues.
First, we can address the robustness of the results across different model
specifications, and conclude if the EKC hypothesis is a robust description
of the income-emission pattern in some country of the region or what other
pattern seems to describe this dynamic. Second, the effect on $CO_{2}$
emissions dynamics of these relevant factors is assessed, providing a better
understanding of the underlying causes of environmental damage. This is a
crucial step to start understanding how to mitigate environmental impact of
growth. The reminder of the paper is structured as follows. Section 2
briefly reviews the theoretical foundations of the hypothesis and highlights
some related criticisms. In Section 3 the econometric methodology as well as
the data employed in the analysis are presented. In Section 4 we present and
discuss the results of the models carried out. Section 5 concludes the paper
and provides some insights for policy recommendations on the basis of our
results.

\section{Theoretical basis of the EKC hypothesis and some related criticisms}

The theory underlying the EKC hypothesis is based on the existence of a
number of time-related effects occurring along the development process.
Indeed, all other things remaining unchanged, greater economic activity
would necessary imply a higher use of resources hence higher environmental
impact. However, this effect, known as scale effect, can be mitigated and
even offset, in the later stages of development, by the dynamic implications
of growth.

A number of underlying factors to the EKC hypothesis have been identified
and different authors have alternatively highlighted the relative importance
of one or the other. Among the direct determinants of the EKC already
identified by Grossman and Krueger (1991), the changes occurring in the
economic structure at different levels of income per capita (i.e. structural
change) could explain a growth-environment inverted U-shaped relationship.
Indeed, as throughout the development process economic structures
traditionally shift from low-polluting agriculture to energy-intensive
industry to lighter manufacture and services, the impact of growth on
environmental quality is expected to change at different levels of income.
This factor has been considered as crucial in explaining the EKC
relationship by many influential authors (Panayotou et al. (2000), for
example) and its relevance has been recently reaffirmed as the
\textquotedblleft first and foremost\textquotedblright\ analytical base of
the EKC (Savona and Ciarli (2019), p. 247). However at least two criticisms
can be directed to the environmentally beneficial impact of structural
change. A first issue is related to the actual level of dematerialization
brought about by a switch to a service economy. Indeed, the idea that the
service sector uses a lesser amount of resources has been questioned (Fix
(2019)). This criticism is related both to the strong interrelation existing
among different sectors of activity, which should be considered as
complements rather than substitutes (Jespersen (1999)) and to the existing
difference between production and consumption patterns. Even if in terms of
production an economy that switches to services can dematerialize, its
consumption patterns will not change accordingly meaning that, when taking a
vertically integrated approach, that is when indirect emissions are
accounted for (\textquotedblleft consumption perspective\textquotedblright
), the overall decrease in environmental pressure related to structural
change towards services is substantially reduced (Marin and Zoboli (2017)).
A second issue related to the composition effect, particularly relevant when
the EKC is applied to explain the income-emissions patterns in developing
countries, is related to the type of structural change the EKC theory refers
to. Indeed, the idea that economies switch from agriculture to industry and
finally services is based on the transformations that occurred in the now
developed countries during the nineteenth and twentieth centuries. However,
in the face of a very different context and of the existence of many
experiences of so-called \textquotedblleft premature
de-industrialization\textquotedblright\ (Palma (2014); Rodrik (2016)) it is
possible to believe that those steps are not being followed by developing
countries in current times, with the related implications in environmental
terms.

Another element essential to the occurrence of environmental quality
improvement as income rises is the technological progress that is generally
associated with development. This factor, that refers both to general
productivity improvements and emission specific changes in process that lead
to an improvement in energy efficiency, has been considered as crucial and
become known as technique effect. It is worth mentioning that the results
reached by most studies of decomposition analysis -- another stream of
literature that seeks to study the income-emissions relationship by
decomposing emissions into their sources of changes (Stern (2017)) -- show
that the within-sector technological change plays the most important role in
explaining energy intensity changes. This conclusion is reached by both
multiple (Voigt et al. (2014) and Jimenez and Mercado (2014) among others)
and single-country (Sinton and Levin (1994); Zhang (2003); Ma and Stern
(2008); Ke et al. (2012) for China and Bhattacharya and Shyamal (2001) for
India) studies. However, while technological progress is so important in
explaining environmental improvement, it is also very unlikely to occur
automatically in developing countries (Zilio (2012)). This may be due to a
number of constraints that range from import of obsolete technologies and
poor own development of new technologies due to low incentives to firms
eco-innovation and meagre public R\&D expenditure.

Along with these more traditional elements explaining the EKC, a number of
additional factors have been taken into account. Input mix changes and
particularly the improvement in the energy mix, which has been found to
occur with income growth (Semieniuk (2018) confirming the \textquotedblleft
energy ladder hypothesis\textquotedblright ), could explain a reduced
environmental impact of production at higher levels of per capita income.
The role of education along with increasing environmental awareness and
changes in consumer preferences have also been identified as underlying
factors to the EKC, given that environmental quality has been considered a
luxury good\footnote{%
The role of these \textquotedblleft behavioral factors\textquotedblright\ in
reducing the environmental impact of production in later stages of
development, strongly emphasized by the early EKC literature, has been
criticized. Panayotou et al. (2000) argue that, since their contribution
requires a perception of the negative impact of increased pollution, they
would be inconsistent with empirical findings of an inverted U-shaped
relationship between growth and global pollutants such as CO2. However, as
Stern points out, these would be \textquotedblright underlying causes [. . .
] which can only have an effect via the proximate
variables\textquotedblright\ (Stern (2004), p. 1421). Still, criticisms
remain given that high income inequality and highly non-linear preferences
for environmentally-friendly goods considerably constrain these products
mass diffusion hence consumer preferences significant impact on the
relationship (Magnani (2000) and Vona and Patriarca (2011)).}. Finally, the
role of the implementation of stricter environmental regulations in more
developed countries has also been highlighted, even if some criticisms have
been raised to this respect. In particular, it has been claimed that even if
stricter environmental regulation could explain the reduction in
environmental damage in some countries, it would hardly support the
existence of an EKC at the global level. Indeed, once stricter regulations
are enforced in one country, firms may relocate their more polluting
activities to countries with laxer rules -- typically developing countries
-- rather than invest in eco-innovation and reduce their total emissions.
This effect, known as the \textquotedblleft pollution-haven
hypothesis\textquotedblright\ (PHH), could be further magnified by trade
liberalization which would reduce the costs of offshoring the
\textquotedblleft dirty\textquotedblright\ production. In fact, this
assumption has been further developed through the \textquotedblleft
pollution offshoring hypothesis\textquotedblright , explicitly linking firms
decisions to relocate highly polluting production to trade liberalization.
These concerns triggered a stream of literature studying the effect of trade
and international relocation of industries on the environment particularly
in developing countries. However, mixed evidence has been found with respect
to the PHH hypothesis, possibly due to the employment of different empirical
approaches.

\section{Methodology and data}

\subsection{Methodology}

A very large stream of empirical literature used different econometric
techniques to test the EKC hypothesis in the last three decades. In
particular, both methodological criticisms directed towards the early
studies ignoring the possible existence of unit root in the data (Stern
(2004)) and the long run nature of the relationship (Dinda (2004)), promoted
the implementation of different univariate and multivariate techniques to
test for long run cointegrating relationships. Among these, the (1) Engle
and Granger (1987) residual based approach to test for cointegrating
relationships; 2) the full information maximum likelihood method developed
by Johansen and Juselius (1990) and 3) the fully modified OLS procedure
developed by Phillips and Hansen (1990) have been used. However, Narayan and
Smith (2005) showed that these tests may be inappropriate when the sample
size is relatively small.

Against this background, in this paper we use Autoregressive Distributed Lag
(hereafter, ARDL) bounds testing approach to cointegration based on
Unrestricted Error Correction Model (hereinafter, UECM) to analyze the long
run relationships. The error correction terms from the UECM are used to test
for the direction of Granger-Causality and to conduct generalized variance
decomposition analysis.

The ARDL bounds test procedure has been extensively used to test the EKC
relationship. For example, using data panel, Fuinhas et al. (2017) used this
procedure to test the impact of renewable energy policies on $CO_{2}$
emissions in a panel of ten LACs and Apergis and Payne (2009) applied this
methodology for Central American countries. Using time series data, this
methodology has been used in Amri (2018) in a study for Tunisia, in B\"{o}l%
\"{u}k and Mert (2015) to test for the EKC relationship controlling for
renewable energy in Turkey, in Onafowora and Owoye (2014) to test the EKC
for several countries including Brazil and Mexico, in Zambrano-Monserrate et
al. (2016) to explore the relationship between carbon dioxide emissions,
economic growth, energy use and hydroelectric electricity production in
Brazil and in Zambrano-Monserrate et al. (2018) testing the EKC hypothesis
in Peru controlling for renewable electricity, petroleum and dry natural gas
consumption. However, none of these studies has considered a large number of
Latin American countries as we do in this paper.

The ARDL method, developed by Pesaran and Shin (2001), has some advantages:

1) The ARDL procedure can be applied to any time series, irrespective of the
order of integration of the variables. That is, the time series can be I(0)
or I(1), so that the uncertainty associated with pretesting the order of
integration is eliminated.

2) This procedure is valid for small samples, avoiding the problem of
asymptotic distributions.

3) The technique can distinguish between dependent and independent variables
and generates estimates for the long run and the short run simultaneously,
eliminating the problem generally associated with omitted variables and
autocorrelation.

The ARDL model can be written as:%
\begin{eqnarray*}
\nabla \ln e_{t} &=&\beta _{0}+\dsum\limits_{i=1}^{m}\beta _{1i}\nabla \ln
e_{t-i}+\dsum\limits_{i=0}^{n}\beta _{2i}\nabla \ln
y_{t-i}+\dsum\limits_{i=0}^{r}\beta _{3i}\nabla \ln (y_{t-i})^{2} \\
&&+\dsum\limits_{i=1}^{s}\beta _{4i}\nabla \ln
(y_{t-i})^{3}+\dsum\limits_{j=5}^{p_{j}}\beta
_{j}\dsum\limits_{i=0}^{r}(\nabla \ln x_{j,t-j}) \\
&&+\delta _{1}\ln e_{t-1}+\delta _{2}\ln y_{t-1}+\delta _{3}\ln
(y_{t-1})^{2}+\delta _{4}\ln (y_{t-1})^{3} \\
&&+\dsum\limits_{j=1}^{p_{j}}\delta _{j}x_{j,t-1}+u_{t}
\end{eqnarray*}%
\bigskip

The lags included in each term of the right hand size are selected using any
Information Criteria. In this expression, the $\beta _{j}$ represent the
short run error correction dynamics, while the terms $\delta _{j}$ $%
(j=1,2,...,p_{j})$ correspond to the long run relationship.

We test for cointegration relationship between the variables in the system,
using the bounds test, developed by Pesaran and Shin (1999, 2001). We use an
F-statistic to detemine whether the variables are cointegrated by testing
the joint significance of the lagged level coefficients; that is: 
\begin{eqnarray*}
H_{0} &:&\delta _{i}=0\;~(i=1,2,..,q)\text{, there is no cointegratiion} \\
H_{1} &:&\delta _{i}\neq 0\text{; there is cointegration}
\end{eqnarray*}%
In the presence of cointegration, one should fail to accept the null
hypothesis.

Narayan (2005) and Narayan and Narayan (2010) derived exact critical values
for the bounds test developped in Pesaran and Shin (2001). They show that
there can be three possible situations. They derived a lower and an upper
bound so that if the F-statitistic is lower than the lower bound, there is
not \ a cointegration relationship. If the F-statistic lies between the
lower and the upper bound, the result of the test is inconclusive and we
should use other techniques to analyze the cointegration. Finally, if the
F-statistic is higher than the upper bound, then we cannot reject that a
cointegration relationship exists.

We use Schwarz information criteria to identify the optimal order of the
ARDL components, that is the logs of the differenced variables (short run
dynamics) Once the optimal lag length are selected and the long run
relationship is confirmed, then the UECM can be estimated:%
\begin{eqnarray*}
\nabla \ln e_{t} &=&\omega _{0}+\dsum\limits_{i=1}^{m}\omega _{1i}\nabla \ln
e_{t-i}+\dsum\limits_{i=0}^{n}\omega _{2,j}\nabla \ln
y_{t-i}+\dsum\limits_{h=0}^{r}\omega _{3h}\nabla \ln (y_{t-i})^{2} \\
&&+\dsum\limits_{k=1}^{s}\omega _{4ik}\nabla \ln
(y_{t-i})^{3}+\dsum\limits_{j=5}^{p_{j}}\omega
_{j}\dsum\limits_{q=0}^{r}(\nabla \ln x_{j,t-q})+\varphi EC_{t-1}+\mu _{t}
\end{eqnarray*}%
where $\varphi $ is the speed of adjustment parameter and $EC_{e-t}$ is the
one period lagged error correction term. This coefficient indicates the
speed of adjustment back to equilibrium after a shock in the system and it
should have statistically significant negative sign.

\subsection{Data}

We used data obtained from the World Bank, CEPALSTAT and Latin American
Energy Organization. The $CO_{2}$ emissions per capita, GDP per capita and
population density data are obtained from the World Bank Development
Indicators. The data on agriculture, industry and services value added to
GDP are also obtained from the World Bank Development Indicators, and the
data on the share of primary products exports are obtained from CEPALSTAT.
GDP per capita, agriculture, industry and services data are expressed in
constant 2010 US dollars. The share of primary exports refers to the share
of total exports of a number of commodities including food, live animals,
mineral fuels, lubricants and related materials. We consider annual data
series for 21 LACs\footnote{%
The LACs considered are: Argentina (ARG), Bolivia (BOL), Brazil (BRA), Chile
(CHI), Colombia (COL), Costa Rica (COS), Cuba (CUBA); Dominican Republic
(R.DOM), Ecuador (ECU), El Salvador (EL SAL), Guatemala (GUA), Haiti (HAI),
Honduras (HON), Jamaica (JAM), Mexico (MEX), Nicaragua (NIC), Panama (PAN),
Paraguay (PAR), Peru (PERU), Uruguay (URU) and Venezuela (VEN).} over the
period 1960-2017\footnote{%
In some of the estimations, the sample period starts in 1970 due to data
availability.}. Other variables included in the analysis are related with
the external relationships of the LACs through trade -- measured by exports
and imports of goods and services -- and foreign direct investment (FDI). We
also consider variables related with the extension of the agricultural
sector in the region (agricultural land and percentage of rural population)
and variables related with energy consumption (electricity, gasoline, diesel
and fuel consumption). A description of the variables, including the related
source and the sample is provided in table 1.

\bigskip

\begin{center}
Insert table 1 around here
\end{center}

\bigskip

\section{Results}

\subsection{Estimation of the basic EKC relationship}

In this first model we estimate the basic relationship between income and $%
CO_{2}$ emissions per capita, including the level and the square of the
income term to test for the EKC using the ARDL specification and the bound
tests. The results of these estimations are shown in table 2. The main
findings are:

-\qquad Only in five out of the twenty-one LACs considered, cointegration
between the variables is found;

-\qquad Of these five countries, only three cases (Costa Rica, Ecuador and
Mexico) show inverted U-shaped relationships supporting the EKC -- that is,
positive parameter of the income in level and negative parameter related to
the squared income term. In two countries (Argentina and Peru) the income
parameters are non-significant at the 5\% level;

-\qquad Only in the case of Haiti the results of the bound test are
inconclusive.

\bigskip

\begin{center}
insert table 2 around here
\end{center}

\bigskip

A first conclusion to be drawn is that only in three out of the twenty-one
countries considered we found results supportive of the EKC. In these cases
the turning points are located inside the sample. In these cases the speed
of adjustment parameter is negative and significant, being around 0.50 in
absolute value. This is an evidence in favor of a cointegration relationship
among the variables ($e_{t}$ and $y_{t},y_{t}^{2}$ ). The speed of
adjustment represents the proportion by which the long run disequilibrium in
the dependent variable is corrected in each short time period. For the other
countries showing a cointegrating relationship, the estimated coefficients
are not significant, so the EKC relationship is not supported by the data.
Finally, for most of the countries no cointegration is found in this
estimation. We can conclude that these estimations do not yield very much
support to the EKC hypothesis. However, some econometric problems, such as
those outlined in M\"{u}ller-F\"{u}rstenberger and Wagner (2007) may be
present in these estimations influencing the results. For example, we might
be missing relevant variables that explain $CO_{2}$ emissions independently
from income per capita, hence suffering a misspecification error. In order
to address this concern and establish if the relationships found are robust
across different specifications we estimate again the relationships
including additional variables in the following ARDL models.

\subsection{Estimation of the EKC relationship controlling for production
structure}

As we have seen, the composition effect, that is the shift of production
structures from agriculture to industry and finally to the service sector
along the development path, has been considered among the most important
causes to the EKC hypothesis. In fact, a greater importance of the
agriculture and services sectors in an economy are expected to be grounds
for less $CO_{2}$ emissions with respect to the industrial sector, typically
considered the most polluting economic activity. However, we also
highlighted some criticisms that have been raised challenging the
composition effect as causing the EKC. On the one hand the actual extent to
which a greater service sector implies a lesser amount of resources used by
the economy has been questioned (Fix (2019); Marin and Zoboli (2017)). On
the other, criticisms with respect to the occurrence of similar structural
change processes in developed countries then and developing countries now
have also been raised. In this respect, it should be considered that Latin
American economies -- that never reached high industrialization levels%
\footnote{%
The occurrence of de-industrialization in many developing countries whose
industrial sector was not fully developed yet has been considered a cause of
concern by different authors and was even referred to as \textquotedblleft
premature de-industrialization\textquotedblright\ (Palma (2014); Rodrik
(2016)).} -- experienced a generalized de-industrialization process since
about the mid-1970s and that since then the service sector has been
increasingly important. Given that industry is the most polluting sector and
that it does not play a central role in LACs' production structure, we might
expect a stronger evidence for the EKC in these countries once output
structure is taken into account. Therefore, in this second model, we include
agriculture, industry and services value added to GDP as additional
explanatory variables. We prefer these variables as proxy for the output
structure and its changes over time against the sectoral contributions to
GDP to minimize eventual collinearity problems among the covariates. The
results of these second estimates are displayed in table 3.

\begin{center}
\bigskip

insert table 3 around here

\bigskip
\end{center}

Compared with the previous model, once output structure is taken into
account we find only three cases of not cointegrating relationships. Both
the number of countries with inconclusive situations and cointegrating
relationships increase. In some of these cases, however, the income
parameters are non-significant, so we cannot draw conclusions on the
existence of support for the EKC for these countries\footnote{%
Argentina and Peru, for which this result is the same as in the previous
model; Ecuador, for which in the first estimation we found support for the
EKC and Cuba, Haiti and Honduras.} . Among the cointegrating relationships
for which the income parameters are significant we find Colombia, Costa
Rica, Jamaica and Mexico. In those countries the signs of the parameters
support the EKC hypothesis. It seems important to note that for Costa Rica
and Mexico we find similar results as in the first estimation, meaning that
those results are likely to be robust, hence describing the real
income-emissions relationship in those countries. The income parameters are
also significant in Venezuela, but the estimated signs point to a U-shaped
relationship in this case.

\subsection{Estimating the EKC controlling for output structure through the
share of primary products exports: investigating the environmental impact of
commodity dependence}

Commodity dependence is a long-time feature of Latin American economies.
With different nuances\footnote{%
Andean economies, which include countries from Venezuela to Chile are mostly
specialized in oil, gas and minerals whereas the rest of South America is
agriculture-based (Ocampo (2017)).} , all LACs' output and export structures
are strongly concentrated in primary products and mostly due to the
well-known boom of commodity prices, this pattern was even exacerbated in
recent years\footnote{%
In 2017 primary products exports accounted on average for 65.1\% of total
exports in the region, according to CEPALSTAT data. We should also note
that, among the LACs in our sample, only Mexico and El Salvador have
partially diversified their export structure away from agriculture over the
study period of 1963-2017. In 2017 the exports of primary products accounted
for 17.9 and 23.9 percent of total exports in these two countries
respectively.} . Surely, commodity dependence has a number of different
implications and its analysis goes far beyond the objectives of this paper.
However, given the great importance of this pattern in the region, it may
have an impact on LACs' emissions dynamics and their relation with income
that is worth considering.

Therefore, we estimate again the model controlling for the export share of
primary products. We choose this variable against the product share of these
goods to minimize the risk of collinearity among the covariates. In this
estimation we also control for population density, which many have
considered as a potential underlying factor to $CO_{2}$ emissions dynamics.
However, there is not complete agreement over the expected impact of this
factor. Some deem increasing population density to reduce, ceteris paribus,
a country's emissions, due to the reduction in transportation and electric
networking costs that it would imply (Panayotou et al., 2000). In contrast,
others have believed that increasing population density increases emissions
given that \textquotedblleft more dense populations will burn more
fuel\textquotedblright . (Poudel et al. 2009, p. 19).

\bigskip

\begin{center}
[insert table 4 around here]

\bigskip
\end{center}

Results in table 4 show the bounds tests and long run estimates of the
cointegrating relationships controlling for population density and exports
of primary goods. In this estimation the number of countries for which we
find cointegration increases again, but only in some of them (Colombia,
Costa Rica, Ecuador, El Salvador, Mexico, Paraguay) we find significant
income parameters. In all these six cases the income parameters signs
provide support to the EKC hypothesis.

With respect to the control variables included in this model we find that
the share of commodity exports parameter, when significant (in Jamaica,
Mexico, Paraguay and Peru) always shows positive sign. This implies that a
higher share of commodity exports is related to a higher environmental
impact as measured by $CO_{2}$ emissions per capita. This result is in line
with the neo-extractivism literature denouncing the high environmental
pressure caused by commodity dependent economies (Lander (2014); Svampa
(2019)). In the light of these results and the fact that a large part of the
environmental impact of this economic model is not captured by $CO_{2}$
emissions dynamics, the analysis in environmental economics perspective of
this production model is an interesting field to be explored by future
research.

Turning to the effect of population density on $CO_{2}$ emissions per capita
our estimations yield mixed results supporting both postures expressed by
the literature, depending on the country. In most of the cases for which the
population density parameter is significant (Cuba, El Salvador, Haiti,
Mexico, Paraguay, Peru) it has positive sign, implying that a higher density
of population -- which is also likely to be associated to higher
urbanization -- increases carbon dioxide emissions per capita. However, in
some cases, for example in Mexico, the estimated sign is negative. In this
country, the large extension of the territory may be explaining this result.
Given its geographical characteristics indeed, it is likely that the
benefits in terms of transportation and networking savings overwhelm the
negative environmental effects of increasing population density.

\subsection{Estimating the EKC controlling for external relationships and
agricultural land}

As already mentioned, the effect of trade on environmental quality has been
extensively discussed, and the idea that trade has a negative impact on the
environment in developing countries has been formulated in the PHH
hypothesis. In order to control for the eventual influence of trade and
other external relationships of LACs, in this model we include FDI inflows,
as well as exports and imports as a share of GDP. Moreover, in this
estimation we also control for the share of agricultural land area, since we
suppose that this variable, reflecting important characteristics of each
country, might be influencing the way their external relations are shaped.

\bigskip

\begin{center}
insert table 5 around here
\end{center}

\bigskip

In this model, the number of cointegrating relationships increases with
respect to the first model. Indeed, we find cointegration for fourteen
countries in the sample, being only three and four respectively the cases
for which either no cointegration or inconclusive results are found.
However, of the countries for which we found cointegration, only eight
(Brazil, Ecuador, El Salvador, Honduras, Jamaica, Mexico, Nicaragua, Panama)
show significant income parameters, and six an inverted U-shaped
relationship (all but Honduras and Nicaragua for which the signs are
indicating a U-shaped relationship). In all the cases where the EKC is
supported, the speed of adjustment in the short run relationship is negative
and clearly statistically significant.

Turning to the analysis of the control variables of this model, we find that
the FDI related parameter is significant in four (Argentina, El Salvador,
Jamaica and Venezuela) out of the fourteen countries for which we find
cointegration. In all these countries except in Jamaica, the sign of the
parameter is positive which might be indicating that FDI are mainly directed
towards polluting sectors, at least in these countries. Indeed,
pollution-intensive sectors attract a large share of total FDI inflows in
the region (Blanco et al. (2013)) and many studies investigating the
environmental impact of FDI found that, in most cases, environmental damage
and pollution are linked to or caused by increasing FDI inflows (Hoffmann et
al. (2005); Merican et al. (2007); Acharyya (2009); Lee (2009)).

In relation to the variables related to trade we find that the export
parameter is significant in only four countries (Honduras, Jamaica, Mexico,
Nicaragua), in most cases showing a negative sign. It seems that a greater
share of exports has a positive environmental effect, if any. With respect
to imports, the parameter is significant in only five countries (Brazil, El
salvador, Jamaica, Mexico, Nicaragua) and shows mixed signs. Overall, we
don't find clear evidence of an univocal environmental impact of trade and
our findings do not support the PHH hypothesis in the region. Rather, it
seems that the environmental impact of trade is different in each country
and it is likely to depend on a variety of issues, ranging from inherent
characteristics of the country to specific regulations implemented. This
result is consistent with the findings of previous literature looking for
the impact of trade in the region and particularly with the conclusions
reached by Jenkins (2003). Analyzing the environmental effect of the
openness to trade after the liberalization process of mid-80s/early 90s in
Argentina, Brazil and Mexico he found that not a unique effect could be
found. In Argentina and Brazil opening to trade resulted in an exacerbation
of their existing specialization in polluting industries. Conversely in
Mexico -- which was the only country among these to implement environmental
regulations together with commercial liberalization -- increasing trade had
beneficial environmental effects. That is, these and our results suggest
that it is likely that the environmental impact of greater commercial
activity is determined by the context within which trade is increased rather
than by trade itself.

Finally, with respect to the parameter related to agricultural land it is
significant in seven countries (Argentina, El Salvador, Honduras, Jamaica,
Mexico, Nicaragua and Venezuela), having positive sign in all of them,
except for Honduras and Mexico. The fact that a greater share of
agricultural land seems to lead to greater emissions can be related to
different factors. It can be related to the fact that greater agricultural
activity is environmentally damaging, differently from what could be
expected and this might be due to the type of agricultural practices carried
out. This is likely to be case of Argentina, for example, in which the
agricultural activity is very much related to an environmentally impacting
agro-industry.

\subsection{Estimating the EKC controlling for renewable energy production
population density and rural population}

Among the many factors that determine the possibility of different
environmental impacts of growth, the energy mix plays an important role. In
particular, when energy is obtained from renewable and clean sources the
impact on the environment is reduced. Fuinhas et al (2017) studied the
effect of renewable energy policies on $CO_{2}$ emissions in ten Latin
American countries and found that, while higher levels of primary energy
consumption per capita lead to higher emissions levels, those can be reduced
in the long run by the implementation of renewable energy policies. Also, a
decomposition analysis by Sheinbaum et al (2011) showed that, despite energy
intensity reductions in Colombia and Mexico -- and to a lesser extent in
Argentina and Brazil --, the increasing dependence on fossil fuels for
energy generation in these countries has hindered a reduction in $CO_{2}$
emissions to occur.

Against this background we include renewable energy production in the model
that estimates the relationship between income and carbon dioxide emissions
per capita to control for the effect of renewable sources of energy on the
environmental impact of growth. In this estimation, we also control for the
independent effect on $CO_{2}$ emissions of population density and the share
of rural population.

The results are displayed in table 6.

\bigskip

\begin{center}
insert table 6 around here
\end{center}

\bigskip

First of all, we note that, as observed for other models, when more control
variables are included the number of countries for which a cointegrating
relationship is found increases. Of the seventeen countries for which we
find cointegration in this estimation, only eleven show significant income
parameters and six (Ecuador, El Salvador, Jamaica, Mexico, Paraguay and
Peru) have signs supporting the EKC hypothesis.

With respect to the control variables included, we observe that the
parameter related to renewable energy production is significant and negative
in most of countries for which cointegration is found. This result is not
surprising considering that renewable energy production generates a lesser
amount of emissions than energy obtained from fossil fuels. However, it is
interesting to note that in Paraguay, where hydroelectric energy generation
is particularly important, the parameter related to renewable energy
production has positive sign, implying that this energy production is
increasing $CO_{2}$ emissions in this country. This issue should be further
investigated.

The population density parameter is significant in twelve countries in this
estimation, with mixed signs. Again, our results support the idea that
higher population density tends to increase emissions in small countries --
for example, Cuba, Dominican Republic El Salvador and Jamaica for which the
parameter has positive sign -- whereas it might have a beneficial impact on
emissions in countries like Chile where more dense populations can
significantly reduce transportation costs and emissions.

Finally, we observe that the share of rural population also seems to have an
impact on emissions per capita. Not surprisingly, a higher share of rural
population is generally leading to lower levels of $CO_{2}$ emissions per
capita -- in most countries in our sample, the parameter is significant and
negative.

\subsection{Estimating the EKC relationship controlling for energy
consumption}

The importance of energy consumption and the different sources of energy
generation has already been discussed. In this model we use the bound tests
and ARDL specification to estimate the long run relationships testing the
EKC hypothesis controlling for electricity, gasoline, diesel and fuel
consumption. The results of these estimations are found in table 7.

\bigskip

\begin{center}
insert table 7 around here
\end{center}

\bigskip

We found cointegration in fourteen out of twenty-one countries in the
sample, and significant income parameters in eight of these countries.
However, according to the signs of the parameters, we only find support for
the EKC in Costa Rica, Cuba, El Salvador and Mexico.

With respect to the variables included to control for energy consumption we
find that they are significant in most countries. Overall, the signs are as
expected. Indeed, the parameters related to diesel, gasoline and fuel
consumption are significant and positive in the vast majority of cases: not
at all surprisingly fossil fuel consumption increases $CO_{2}$ emissions per
capita. Conversely, mixed results are found for the electricity consumption
parameter that is positive or negative depending on the country. This is
likely to depend on the source of electricity generation in each country as
well as on the extent to which electricity is substituting energy
consumption from other more or less environmentally damaging sources.

\subsection{Discussion of the results}

The results of the estimates performed point out to mixed results. We find
that the number of countries for which we find cointegration in the
different model specifications, varies depending on the variables we control
for. When the model includes more control variables we observe an increase
in the number of countries that show cointegrating relationships. In the
cases in which we find cointegration, not always the income parameters are
significant for all countries. As a consequence, in some cases it is not
possible to define which pattern carbon dioxide emissions follow as income
grows. Moreover, even when the income parameters are significant, not always
they have the signs predicted by the EKC hypothesis. Overall, our results do
not support the EKC hypothesis for most countries in the region, implying
that we cannot expect an automatic reduction of $CO_{2}$ emissions per
capita with income growth, even in the long term. However, there is a
minority of countries for which we find fairly consistent results supporting
the EKC hypothesis. In the case of Mexico, we find support for the EKC in
all the six models performed and the estimated turning points are also
stable, ranging from a minimum of 8993 US\$ in model 4 to a maximum of
11312.9 US\$ in model 6. The turning point estimated is located inside the
sample in models shown in tables 2, 3, 5 and 6. Similar results are obtained
for Costa Rica and El Salvador. In these cases we find support for the EKC
in four out of the six models performed and the values estimated for the
income parameters also are quite robust resulting in pretty stable turning
point estimates -- at about 9000 US\$ and 2800 US\$ in Costa Rica and El
Salvador respectively. In Ecuador, the EKC hypothesis is confirmed in four
out of six cases as well, but the estimates for the turning points are less
robust, ranging from 4498 to 9059 US\$. In\ the case of Casta Rica, the
turning points are inside the sample in models shown in tables 4 and 7. The
same happens in the case of El Salvador for models shown in tables 4 to 7.
Conversely in the case of Ecuador, the turning points are inside the sample
in models shown in tables 3 and 5. In other countries we also observe
results confirming the EKC hypothesis, but those results are not robust
across different model specifications. Overall, we can conclude that in most
of the countries of the region the EKC is not supported by evidence.
However, there are some countries -- namely Mexico, Costa Rica, El Salvador
and Ecuador -- for which the relationship between $CO_{2}$ emissions and
income per capita seems to be described by an inverted U-shaped curve in the
long term.

With respect to the control variables included in the different models we
observe that these are in general significant meaning that not considering
them might create problems of omitted variables bias. Even if not all
control variables are significant in all the countries considered and even
if in some cases the results are mixed, the signs of the parameters are as
expected in most cases. This allows us to draw some conclusions about the
environmental impact of some elements related to the process of development
that have been considered as either causing environmental damage or allowing
environmental improvement, independently from income growth.

With respect to the investigation of the composition effect in LACs, our
results do not provide evidence of a unique effect of the environmental
impact of the industrial and service sectors. However, we find that the
primarization of LACs' economies tends to have a negative impact on the
environment, as measured by $CO_{2}$ emissions. This conclusion is supported
by the observation that both the share of commodity exports and the share of
agricultural land tend to increase $CO_{2}$ emissions per capita.
Considering that a higher share of rural population is found to reduce $%
CO_{2}$ emissions per capita, we consider this result as related to the
commodities production model in the region rather than to the sector itself.
In this sense, the primary sector that could be modestly impactful on the
environment, ends up exerting high environmental damage due to the way it is
deployed in the region -- in the form of mining activities or agroindustry.

Among the factors that are also found to be relevant in determining the
environmental impact of growth we find the energy mix to play an important
role. Indeed, if on the one hand a higher consumption of fossil fuel
produced energy tends to increase $CO_{2}$ emissions, they are reduced if a
higher share of renewable energy is produced. As a consequence, the
environmental effect of economic growth is not only determined by the level
of such growth, but the way the additional income is produced is also
extremely important.

With respect to other elements considered in our analysis, we find more
mixed results. Indeed, the environmental effect of both population density
and external relationships seems to highly depend on the individual country
considered. As a general consideration, we might say that population density
tends to reduce $CO_{2}$ emissions only in those countries that have a
geographical configuration that makes the benefits of more dense population
-- in terms of networking and reduction of transportation costs --
particularly important. Moreover, with respect to the effect of FDI inflows
and trade related variables, we observed that not a common pattern exists.
Our results provide some evidence that FDI inflows tend to increase $CO_{2}$
emissions in most countries of the region, but no clear support for the PHH
is provided by our results.

\section{Conclusions}

Based on the influential Environmental Kuznets Curve hypothesis, in this
paper we employed an ARDL bounds testing approach to cointegration and
Unrestricted Error Correction Models to estimate the relationship between
income and $CO_{2}$ emissions per capita in twenty-one Latin American
Countries over 1960-2017. Following a time series approach we performed a
separate estimation for each one of the countries in our sample. We
estimated six different specifications of the model for each country, to
take into account the independent effect on $CO_{2}$ emissions per capita of
different factors other than income. The analysis performed in this paper
was aimed at addressing two specific concerns. First, through the estimation
of the model controlling for different variables, we wanted to assess if the
EKC hypothesis was confirmed in any of the LACs considered, and if its
validity was robust across different model specifications, given that the
high volatility of the estimates is one of the main criticisms concerning
the EKC hypothesis. Second, the inclusion of control variables accounting
for the effect on $CO_{2}$ emissions of a number of relevant factors, should
serve to increase our understanding of the causes of change of emissions in
different countries, which is an important premise to design effective
mitigation policies.

With respect to the first point, we might say that the EKC hypothesis is not
supported by evidence in most countries in the region. However, we also
found that in a minority of countries in our sample the EKC hypothesis seems
to effectively describe the income-emissions relationship in the long term.
Indeed, we find robust results supporting the EKC hypothesis in Mexico,
Costa Rica, El Salvador and Ecuador. The finding that different patterns for
the dynamics of $CO_{2}$ emissions at different levels of income per capita
apply to different countries in the region enhances the importance of
analyzing this relationship at the country level. Indeed, there is not a
unique pattern that describes this relationship for all countries even in
the same region, and individual experiences are extremely important in
defining the dynamics of $CO_{2}$ emissions per capita along the development
path. In this sense, we observed that the inherent geographical
characteristics of the country and the policies implemented, particularly
environmental regulations, are likely to explain different effects on $%
CO_{2} $ emissions of population density and trade, respectively. On the
other hand, we also observed that some common patterns in the region,
particularly commodity dependence and the specific production models of
these primarized economies, exert similar negative effects on environmental
quality. In this respect, the promotion of a production and export structure
less concentrated in commodities is advisable in the region. This virtuous
structural change, if pursued, could provide these countries with better
environmental performance, in addition to more stable growth. Moreover, the
development of a greener energy mix, particularly through the fostering of
renewable energy production, is advisable in consideration of the reduced
environmental impact of energy consumption from these sources, against
fossil fuel produced energy.

In conclusion, both the observation that the EKC hypothesis describes the
income-emissions pattern only in a minority of countries in the region and
the manifest importance of some country-specific policies in determining a
lesser environmental impact of growth, call for environmental policy in the
region. These policies, in the form of promotion of renewable energy
production, fostering of greener sectors of economic activity or
environmental standards to be implemented in conjunction to increased
international openness, can strongly influence the impact of growth on $%
CO_{2}$ emissions in these countries allowing them to capture the economic
benefits of it without exerting ever increasing environmental damage.

\bigskip

\section{References}

Acharyya, J. (2009) FDI, growth and the environment: evidence from India on $%
CO_{2}$ emission during the last two decades, Journal of Economic
Development, 34(1), pp. 43--58.

Amri, F\textbf{. }\ (2018). \ Carbon dioxide emissions, total factor
productivity, ICT, trade, financial development, and energy consumption:
testing environmental Kuznets curve hypothesis for Tunisia. Environmental
Science and Pollution Research, 25, pp. 33691-33701.

Aspergis, N. and Payne, J.E\textbf{. }(2009). $CO_{2}$ emissions, energy
usage and output in Central America. Energy Policy, 37, pp. 3282-3286.

Bhattacharya, R. and Shyamal, P., (2001). Sectoral changes in consumption
and intensity energy in India. Indian Economic Review 36 (2), 381-392.

Blanco, L. , Gonza\~{n}ez, F. and Ruiz, I.\textbf{\ }(2013). The impact of
FDI on $CO_{2}$ emissions in Latin America. Oxford Development Studies, vol.
41(1), pp. \ 104-121.

B\"{o}l\"{u}k, G. and Mert, M\textbf{. }(2015). The renewable energy, growth
and environmental Kuznets curve in -Tukey: An ARDL approach. Renewable and
Sustainable Energy Reviews, 52, pp. 587-595.

Brundtland, G., \textbf{(}1987). Report of the World Commission on
Environment and Development: Our Common Future. United Nations General
Assembly document A/42/427.

de Bruyn, S.M., van den Bergh, J.C.J.M. and Opschoor, J.B., (1998). Economic
growth and emissions: reconsidering empirical basis of environmental Kuznets
curves. Ecological Economics 25 (2), 161-175 (May).

Dinda, S. (2004). Environmental Kuznets curve hypothesis: a survey.
Ecological economics, 49(4), 431-455.

Ehrlich, P.and Holden, J., (1971). Impact of population growth. Science 171,
1212-17 (Mar).

Engle, R.F. and Granger, C.W.J, (1987), Co-Integration and Error Correction:
Representation, estrimation and testing. Econometrica, 55(2), 251-276.

Fix, B. (2019). Dematerialization through services: Evaluating the evidence.
BioPhysical Economics and Resource Quality, 4(2), 6.

Fuinhas, J.A., Cardoso Marques, A. and Koenghan, M. (2017). Are renewable
energy policies upsetting carbon dioxide emissions? The case of Latin
America countries. Environ. Sci. Pollut. Res. 24, pp. 15044 - 15054.

Grossman, G.M. and Krueger, A.B., (1991). Environmental impacts of the North
American Free Trade Agreement. NBER. Working paper 3914.

Hausmann, R., Pritchett, L. and Rodrik, D., (2005). Growth Accelerations.
Journal of Economic Growth 10 (4), 303-329.

Hoffmann, R., Lee, C. -G., Ramasamy, B. \& Yeung, M. (2005) FDI and
pollution: a Granger causality test using panel data, Journal of
International Development, 17(3), pp. 311--317.

Holtz-Eakin, D. and Selden, T.M., (1995). Stoking the fires. $CO_{2}$
emissions and economic growth. Journal of Public Economics 57, 85-101.

Jenkins, R. O. (2003). La apertura comercial \textquestiondown ha creado para%
\'{\i}sos de contaminadores en Am\'{e}rica Latina?. Revista de la CEPAL.

Jespersen, J. (1999). Reconciling environment and employment by switching
from goods to services? A review of Danish experience. European Environment,
9(1), 17-23.

Jimenez, R. and Mercado, J., (2014). Energy intensity: a decomposition and
counterfactual exercise for Latin American countries. Energy Economics. 42,
161-171 (Mar).

Johansen, S. ,1988. "Statistical analysis of co-integrating vectors".
Journal of Economic Dynamics and Control, 12, pp. 231--254.

Johansen, S., and Juselius, K. ,(1990). "Maximum likelihood estimation and
inference on co-integration with application to the demand for money".Oxford
Bulletin of Economics and Statistics, 52, pp.169--221

Ke, J., Price, L., Ohshita, S., Fridley, D., Zheng, N., Zhou, N. and Levine,
M., (2012). China's industrial energy consumption trends and impacts of the
top-1000 enterprises energy-saving program and the ten key energy-saving
projects. Energy Policy 50, 562-569.

Kuznets, S., (1955). Economic growth and income inequality. American
Economic Review, 49, 1-28.

Lander, E. (2014). El Neoextractivismo como modelo de desarrollo en Am\'{e}%
rica Latina y sus contradicciones. Heirinch Boll Stiftung, Berlin.

Lee, C. G. (2009) Foreign direct investment, pollution and economic growth:
evidence from Malaysia, Applied Economics, 41(13), pp. 1709--1716.

Lindmark, M. (2002). An EKC-pattern in historical perspective: carbon
dioxide emissions, technology, fuel prices and growth in Sweden 1870--1997.
Ecological economics, 42(1-2), 333-347.

Ma, C., Stern, D.I., 2008. China's changing energy intensity trend: a
decomposition analysis. Energy Economics 30, 1037-1053.

Magnani, E., (2000). The environmental Kuznets curve, environmental
protection policy and income distribution. Ecological Economics, 32, 431-443.

Marin, G., \& Zoboli, R. (2017). The Economic And Environmental Footprint Of
The Eu Economy: Global Effects Of A Transition To Services. Rivista
Internazionale di Scienze Sociali, 136(2), 195-228.

Martinez-Zarzoso, I., \& Bengochea-Morancho, A. (2003). Testing for an
environmental Kuznets curve in Latin-American countries. Revista de An\'{a}%
lisis Econ\'{o}mico, 18(1), 3-26.

Meadows, D. H., Meadows, D. L., Randers, J. and Behrens W. H., 1972. The
limits to growth; a Report to the Club of Rome's project on the predicament
of mankind. Universe Books, New York.

Merican, Y., Zulkronian, Y., Zaleha, M. \& Hook, L. S. (2007) Foreign direct
investment and the pollution in five ASEAN nations, International Journal of
Economics and Management, 1(2), pp. 245--261.

Moomaw, W.R. and Unruh, G.C., (1997). Are environmental Kuznets curves
misleading us? The case of $CO_{2}$ emissions. Environment and Development
Economics 2 (4), 451-463 (Nov).

Moomaw, W.R. and Unruh, G.C., (1998). An Alternative Analysis of apparent
EKC-type Transitions. Ecological Economics 25 (2), 221-229 (May).

M\"{u}ller-F\"{u}rstenberger, G. and Wagnar, M. (2007). Exploring the
environmental Kuznets hyphotesis: theoretical and econometric problems.
Ecological Economics, 62, pp. 648-660.

Narayan, P.K. (2005) The saving and investmemt nexus for China: evidence
from cointegration tests.\ Applied Econometrics, 37, pp. 1979-1990.

Narayan, P.K and Narayan, S. (2010). Carbon dioxide emissions and economic
growth: panel data evidence from developing countries. Energy Policy, 38(1),
pp. 661-666.

Narayan, P.K, and Smyth, R. (2008). Energy consumption and real GDP in G7
countries: New evidence from panel cointegration with structural breaks.\
Energy Economics, 30, pp. 2331-2341.

Nordhaus, W.D., (1977). Economic growth and climate: the carbon dioxide
problem. American Economic Review, 67 (1), 341-346.

Ocampo, J.A., (2017). Commodity-Led Development in Latin America.
International Development Policy, Revue internationale de politique de d\'{e}%
veloppement 9 (1), 51-76.

Onafowora, O.A. and Owoye, O. (2014). Bounds testing approach to analysis of
the environmental Kuznets. Energy Economics, 44. pp. 47-62.

Palma, J.G., (2014). De-industrialisation, `premature' de-industrialisation
and the dutch-disease. Revista NECAT 5 (1), 7-23.

Panayotou, T., Peterson, A. and Sachs, J.D., (2000). Is the Environmental
Kuznets Curve Driven by Structural Change? What Extended Time Series May
Imply for Developing Countries.

Pesaran, M.H. and Shin, Y. (1999).An autoregressive distributed lag
modelling approach to cointegration analysis. Chapter 11 in Econometrics and
Economic Theroy in the 20th century: The Ragnar Fritch Centenial Symposium.
Stoom S. (ed). Cambridge University Press. Cambridge.

Pesaran, M.H., Shin, Y. and Smyth, R.J. (2001). Bounds testing approaches to
the analysis of level relationships. Journal of Applied Econometrics, 16,
pp. 289-326.

Phillips, P.C.B. and Hansen, B.E. (1990). Statistical inference in
instrumental variables regression with I(1) processes. Review of Economic
Studies, 57, pp. 99-125.

Poudel, B.N., Paudel, K.P. and Bhattarai, K., (2009). Searching for an
Environmental Kuznets Curve in Carbon Dioxide Pollutant in Latin American
Countries. Journal of Agricultural and Applied Economics 41 (1), 13-27 (Apr).

Rodrik, D., (2016). Premature deindustrialization. Journal of Economic
Growth 21 (1), 1-33.

S\'{a}nchez, L., \& Caballero, K. (2019). La curva de Kuznets ambiental y su
relaci\'{o}n con el cambio clim\'{a}tico en Am\'{e}rica Latina y el Caribe:
un an\'{a}lisis de cointegraci\'{o}n con panel, 1980-2015. Revista de econom%
\'{\i}a del Rosario, 22(1), 101-142.

Savona, M. and Ciarli, T., (2019). Structural Changes and Sustainability. A
Selected Review of the Empirical Evidence. Ecological Economics 159, 244-260
(May).

Selden, T.M. and Song, D., (1994). Environmental quality and development: Is
there a Kuznets curve for air pollution? Journal of Environmental Economics
and Management 27, 147--162.

Semieniuk, G., (2018). Energy in Economic Growth: Is Faster Growth Greener?
In: SOAS Department of Economics Working Papers 208. School of Oriental and
African Studies, University of London, London.

Shafik, N., 1994. Economic development and environmental quality: An
econometric analysis. Oxford Economic Papers 46, 757--773.

Shafik, N. and Bandyopadhyay, S., (1992). Economic Growth and Environmental
Quality: Time Series and Crosscountry Evidence. Background Paper for the
World Development Report 1992, The World Bank, Washington DC.

Sheinbaum, C., Ruiz, B.J. and Ozawa, L. (2011). Energy consumption and
related $CO_{2}$ emissions in five Latin American countries: Changes from
1990 to 2006 and perspectives. Energy, 36, pp. 3629-3638.

Sinton, J.E. and Levine, M.D., (1994). Changing energy intensity in Chinese
industry: the relatively importance of structural shift and intensity
change. Energy Policy 22 (3), 239-255.

Stern, D.I., (2004). The rise and fall of the environmental Kuznets curve.
World Development, 32 (8), 1419-1439.

Stern, D.I. and Common, M.S., (2001). Is there an environmental Kuznets
curve for sulfur? Journal of Environmental Economics and Management 41,
162-178.

Stern, D.I., Common, M.S. and Barbier, E.B., (1996). Economic Growth and
Environmental Degradation: the Environmental Kuznets Curve and Sustainable
Development. World Development 24 (7), 1151-1160 (Jul).

Stern, D.I. and Enflo, K., (2013). Causality between energy and output in
the long-run. Energy Economics 39, 135-146.

Svampa, M. (2019). Las fronteras del neoextractivismo en Am\'{e}rica Latina:
conflictos socioambientales, giro ecoterritorial y nuevas dependencias.
Transcript Verlag. CALAS - Maria Sibylla Merian Center.

Unruh, G. C., \& Moomaw, W. R. (1998). An alternative analysis of apparent
EKC-type transitions. Ecological Economics, 25(2), 221-229.

Vincent, J., (1997). Testing for Environmental Kuznets Curves Within a
Developing Country. Environment and Development Economics 2 (4), 417-431
(Nov).

Voigt, S., De Cian, E., Schymura, M. and Verdolini, E., (2014). Energy
intensity developments in 40 major economies: structural change or
technology improvement? Energy Economics 41, 47-62 (Jan).

Vona, F. and Patriarca, F., (2011). Income inequality and the development of
environmental technologies. Ecological Economics 70 (11), 2201-2213 (Sep).

Zambrano-Monserrate, M.A., Valverde-Baja\~{n}a, I., Aguilar-Bohorquez, J.
and Mendoza-Jim\~{n}enez, M. (2016). Relationship between economic growth
and environmental degradation: Is there an environmental evidence of Kuznets
curve for Brazil?. International Journal of Energy Economics and Policy,
6(2), pp. 208-216.

Zambrano-Monserrate, M.A., \ Silva-Zambrano, C., Davalos-Pe\~{n}afiel, J.L.,
Zambrano-Monserrate, A. and Ruano, M.A. (2018). Testing environmental
Kuznets curve hypothesis in Per\'{u}: The role of renewable electricity,
petroleum and dry natural gas.\ Renewable and Sustainable Energy Reviews.,
82, pp. 4170-4179.

Zhang, Z.X., (2003). Why did the energy intensity fall in China's industrial
sector in the 1990s? The relative importance of structural change and
intensity change. Energy Economics 25 (6), 625-638.

Zilio, M. I. (2012). Curva de Kuznets ambiental: La validez de sus
fundamentos en pa\'{\i}ses en desarrollo. Cuadernos de econom\'{\i}a,
35(97), 43-54.

Zilio, M., \& Caraballo, M. (2014). \textquestiondown El final de la curva
de Kuznets de carbono? Un an\'{a}lisis semiparam\'{e}trico para la Am\'{e}%
rica Latina y el Caribe. El trimestre econ\'{o}mico, 81(321), 241-270.

\newpage

\textbf{Tables:}

Table 1: Data employed in the analysis. Definition, sample periods and
sources.

\hspace{-4cm}%
\begin{tabular}{llll}
Variable & Definition & Source & Sample \\ \hline
$e_{t}$ & $CO_{2}$ per capita emissions - tonnes per capita & World Bank
Development Indicators & 1960 - 2017 \\ 
$y_{t}$ & $GDP$ per capita - 2010 Constant \$ & World Bank Development
Indicators & 1960 - 2017 \\ 
$x_{1,t}$ & Agriculture value added to GDP- Constant 2010 US \$ & World Bank
Development Indicators & 1970 - 2017 \\ 
$x_{2,t}$ & Industry value added to GDP- Constant 2010 US \$ & World Bank
Development Indicators & 1970 - 2017 \\ 
$x_{3,t}$ & Services value added to GDP- Constant 2010 US \$ & World Bank
Development Indicators & 1970 - 2017 \\ 
$x_{4,t}$ & Exports of primary goods as a share of total exports & CEPAL -
CEPALSTAT & 1963 - 2017 \\ 
$x_{5,t}$ & Population Density Persons by Km$^{2}$ & World Bank Development
Indicators & 1960 - 2017 \\ 
$x_{6,t}$ & Foreign Direct Investment - net inflows -\% of GDP & World Bank
Development Indicators & 1970 - 2018 \\ 
$x_{7,t}$ & Exports of good and services - \% of GDP & World Bank
Development Indicators & 1960 - 2018 \\ 
$x_{8,t}$ & Imports of good and services - \% of GDP & World Bank
Development Indicators & 1960 - 2018 \\ 
$x_{9,t}$ & Agricultural land - \% of land area & World Bank Development
Indicators & 1970 - 2018 \\ 
$x_{10,t}$ & Renewal Electricity Production - per capita & CEPAL - CEPALSTAT
& 1970 - 2017 \\ 
$x_{11,t}$ & \% of Rural Population & Latin American Energy Organization & 
1970 - 2018 \\ 
$x_{12,t}$ & Diesel oil consumption - per capita & Latin American Energy
Organization & 1970 - 2018 \\ 
$x_{13,t}$ & Electricity consumption - per capita & Latin American Energy
Organization & 1970 - 2018 \\ 
$x_{13,t}$ & Gasoline oil consumption - per capita & Latin American Energy
Organization & 1970 - 2018 \\ 
$x_{14,t}$ & Fuel oil consumption - per capita & Latin American Energy
Organization & 1970 - 2018 \\ \hline
\end{tabular}%
\bigskip

\newpage

\bigskip

Table 2: Results of the bounds test and estimation of the long-run
relationship for the cuadratic EKC relationship $e_{t}=\beta _{0}+\beta
_{1}y_{t}+\beta _{2}y_{t}{}^{2}+u_{t}$

\hspace{-4cm}%
\begin{tabular}{llllccclc}
Country & ARDL model & Bounds test & Conclusion & $\beta _{0}$ & $\beta _{1}$
& $\beta _{2}$ & Turning point & $EC_{t-1}$ \\ \hline
$ARG$ & \multicolumn{1}{c}{$(2,0,0)$} & \multicolumn{1}{c}{$9.802$} & 
\multicolumn{1}{c}{$CI$} & $33.64$ & $-7.777$ & $0.665$ & \multicolumn{1}{c}{%
$4282.96$} & $-0.7396^{a)}$ \\ 
$BOL$ & \multicolumn{1}{c}{$(1,0,1)$} & \multicolumn{1}{c}{$1.8802$} & 
\multicolumn{1}{c}{$NOT~CI$} &  &  &  & \multicolumn{1}{c}{} &  \\ 
$BRA$ & \multicolumn{1}{c}{$(1,0,1)$} & \multicolumn{1}{c}{$1.5295$} & 
\multicolumn{1}{c}{$NOT~CI$} &  &  &  & \multicolumn{1}{c}{} &  \\ 
$CHI$ & \multicolumn{1}{c}{$(1,0,1)$} & \multicolumn{1}{c}{$1.2032$} & 
\multicolumn{1}{c}{$NOT~CI$} &  &  &  & \multicolumn{1}{c}{} &  \\ 
$COL$ & \multicolumn{1}{c}{$(1,1,0)$} & \multicolumn{1}{c}{$1.5022$} & 
\multicolumn{1}{c}{$NOT~CI$} &  &  &  & \multicolumn{1}{c}{} &  \\ 
$COS$ & \multicolumn{1}{c}{$(1,3,0)$} & \multicolumn{1}{c}{$7.0367$} & 
\multicolumn{1}{c}{$CI$} & $-87.567^{a)}$ & $19.277^{a)}$ & $-1.0556^{a)}$ & 
\multicolumn{1}{c}{$9235.41$} & $-0.6301^{a)}$ \\ 
$R.DOM$ & \multicolumn{1}{c}{$(1,1,0)$} & \multicolumn{1}{c}{$2.9504$} & 
\multicolumn{1}{c}{$NOT~CI$} &  &  &  & \multicolumn{1}{c}{} &  \\ 
$CUBA$ & \multicolumn{1}{c}{$(3,0,0)$} & \multicolumn{1}{c}{$1.3233$} & 
\multicolumn{1}{c}{$NOT\ CI$} &  &  &  & \multicolumn{1}{c}{} &  \\ 
$ECU$ & \multicolumn{1}{c}{$(3,0,0)$} & \multicolumn{1}{c}{$5.7418$} & 
\multicolumn{1}{c}{$CI$} & $-254.90^{a)}$ & $60.048^{a)}$ & $-3.524^{a)}$ & 
\multicolumn{1}{c}{$5013.55$} & $-0.4492^{a)}$ \\ 
$ELSAL$ & \multicolumn{1}{c}{$(2,2,0)$} & \multicolumn{1}{c}{$3.336$} & 
\multicolumn{1}{c}{$NOT~CI$} &  &  &  & \multicolumn{1}{c}{} &  \\ 
$GUA$ & \multicolumn{1}{c}{$(1,0,1)$} & \multicolumn{1}{c}{$1.0825$} & 
\multicolumn{1}{c}{$NOT\ CI$} &  &  &  & \multicolumn{1}{c}{} &  \\ 
$HAI$ & \multicolumn{1}{c}{$(1,1,1)$} & \multicolumn{1}{c}{$3.8588$} & 
\multicolumn{1}{c}{$Inconclusive$} &  &  &  & \multicolumn{1}{c}{} &  \\ 
$HON$ & \multicolumn{1}{c}{$(1,0,0)$} & \multicolumn{1}{c}{$0.7988$} & 
\multicolumn{1}{c}{$NOT\ CI$} &  &  &  & \multicolumn{1}{c}{} &  \\ 
$JAM$ & \multicolumn{1}{c}{$(3,0,1)$} & \multicolumn{1}{c}{$1.6462$} & 
\multicolumn{1}{c}{$NOT~CI$} &  &  &  & \multicolumn{1}{c}{} &  \\ 
$MEX$ & \multicolumn{1}{c}{$(1,0,0)$} & \multicolumn{1}{c}{$8.3755$} & 
\multicolumn{1}{c}{$CI$} & $-185.57^{a)}$ & $41.005^{a)}$ & $-2.2446^{a)}$ & 
\multicolumn{1}{c}{$9265.93$} & $-0.5117^{a)}$ \\ 
$NIC$ & \multicolumn{1}{c}{$(1,1,0)$} & \multicolumn{1}{c}{$2.7705$} & 
\multicolumn{1}{c}{$NOT\ CI$} &  &  &  & \multicolumn{1}{c}{} &  \\ 
$PAN$ & \multicolumn{1}{c}{$(1,0,0)$} & \multicolumn{1}{c}{$2.0131$} & 
\multicolumn{1}{c}{$NOT\ CI$} &  &  &  & \multicolumn{1}{c}{} &  \\ 
$PAR$ & \multicolumn{1}{c}{$(1,0,0)$} & \multicolumn{1}{c}{$2.6585$} & 
\multicolumn{1}{c}{$NOT\ CI$} &  &  &  & \multicolumn{1}{c}{} &  \\ 
$PERU$ & \multicolumn{1}{c}{$(1,0,0)$} & \multicolumn{1}{c}{$8.8363$} & 
\multicolumn{1}{c}{$CI$} & $-37.453^{c)}$ & $8.0952$ & $-0.4265$ & 
\multicolumn{1}{c}{$13230.76$} & $-0.5212^{a)}$ \\ 
$URU$ & \multicolumn{1}{c}{$(3,0,2)$} & \multicolumn{1}{c}{$3.0064$} & 
\multicolumn{1}{c}{$NOT\ CI$} &  &  &  & \multicolumn{1}{c}{} &  \\ 
$VEN$ & \multicolumn{1}{c}{$(1,0,0)$} & \multicolumn{1}{c}{$3.0397$} & 
\multicolumn{1}{c}{$NOT\ CI$} &  &  &  & \multicolumn{1}{c}{} &  \\ \hline
\end{tabular}%
\bigskip

\hspace{-4cm}

Narayan (2005)'s bounds test critical values: 1\%: lower bound = 4.8; 1\%
upper bound = 5.725; 5\% lower bound =\ 3.368; 5\% upper bound =\ 4.205.

$^{a)}$significant parameter at 1\%; $^{b)}$ significant parameter at 5\%; $%
^{c)}$ significant parameter at 10\%; $e_{t}=\ln (CO_{2,pc});y_{t}=\ln
(GDP);y_{t}^{2}=(\ln (GDP))^{2};EC_{t-1}:$ estimation of the cointegration
error in the short run relationship or speed of the adjustment in the UECM
for $\nabla e_{t}$\bigskip

\bigskip

\bigskip \newpage

Table 3: Results of the bounds test and estimation of the long-run
relationship for the cuadratic EKC relationship controlling for output
structure $e_{t}=\beta _{0}+\beta _{1}y_{t}+\beta _{2}y_{t}{}^{2}+\beta
_{3}x_{1,t}+\beta _{4}x_{2,t}+\beta _{5}x_{3,t}+u_{t}$

\hspace{-3cm}%
\begin{tabular}{llllcccccclc}
Country & ARDL model & Bounds test & Conclusion & $\beta _{0}$ & $\beta _{1}$
& $\beta _{2}$ & $\beta _{3}$ & $\beta _{4}$ & $\beta _{5}$ & Turning point
& $EC_{t-1}$ \\ \hline
$ARG$ & \multicolumn{1}{c}{$(1,0,0,0,0,0)$} & \multicolumn{1}{c}{$7.195$} & 
\multicolumn{1}{c}{$Ci$} & $27.592$ & $8.700$ & $0.488$ & $0.132$ & $%
0.680^{b)}$ & $-0.299^{a)}$ & \multicolumn{1}{c}{$7373.15$} & $-0.587^{a)}$
\\ 
$BOL$ & \multicolumn{1}{c}{$(1,0,0,0,0,1)$} & \multicolumn{1}{c}{$3.435$} & 
\multicolumn{1}{c}{$Inconclusive$} &  &  &  &  &  &  & \multicolumn{1}{c}{}
&  \\ 
$BRA$ & \multicolumn{1}{c}{$(1,0,1,0,0,0)$} & \multicolumn{1}{c}{$1.308$} & 
\multicolumn{1}{c}{$NOT\ CI$} &  &  &  &  &  &  & \multicolumn{1}{c}{} &  \\ 
$CHI$ & \multicolumn{1}{c}{$(1,0,0,0,0,1)$} & \multicolumn{1}{c}{$3.454$} & 
\multicolumn{1}{c}{$Inconclusive$} &  &  &  &  &  &  & \multicolumn{1}{c}{}
&  \\ 
$COL$ & \multicolumn{1}{c}{$(1,0,0,0,0,0)$} & \multicolumn{1}{c}{$4.113$} & 
\multicolumn{1}{c}{$CI$} & $-49.40^{a)}$ & $16.50^{a)}$ & $-0.823^{a)}$ & $%
-0.410$ & $0.066$ & $-0.910^{a)}$ & \multicolumn{1}{c}{$22514.2$} & $%
-0.480^{a)}$ \\ 
$COS$ & \multicolumn{1}{c}{$(2,0,0,0,0,0)$} & \multicolumn{1}{c}{$4.976$} & 
\multicolumn{1}{c}{$CI$} & $-161.0^{a)}$ & $37.184^{a)}$ & $-2.038^{a)}$ & $%
0.323^{b)}$ & $-0.511$ & $-0.140$ & \multicolumn{1}{c}{$9197.62$} & $%
-0.900^{a)}$ \\ 
$CUBA$ & \multicolumn{1}{c}{$(2,0,0,0,3,0)$} & \multicolumn{1}{c}{$5.308$} & 
\multicolumn{1}{c}{$CI$} & $25.51$ & $10.533$ & $-0.377$ & $-0.271$ & $%
-1.056 $ & $-2.307^{a)}$ & \multicolumn{1}{c}{$116589.5$} & $-0.428^{a)}$ \\ 
$R.DOM$ & \multicolumn{1}{c}{$(1,1,0,0,0,0)$} & \multicolumn{1}{c}{$3.388$}
& \multicolumn{1}{c}{$Inconclusive$} &  &  &  &  &  &  & \multicolumn{1}{c}{}
&  \\ 
$ECU$ & \multicolumn{1}{c}{$(2,0,0,1,0,0)$} & \multicolumn{1}{c}{$8.115$} & 
\multicolumn{1}{c}{$CI$} & $-19.789$ & $-0.515$ & $0.033$ & $-1.032^{a)}$ & $%
0.417$ & $1.673^{a)}$ & \multicolumn{1}{c}{$2462.91$} & $-1.389^{a)}$ \\ 
$ELSAL$ & \multicolumn{1}{c}{$(2,1,0,0,0,0)$} & \multicolumn{1}{c}{$3.207$}
& \multicolumn{1}{c}{$Inconclusive$} &  &  &  &  &  &  & \multicolumn{1}{c}{}
&  \\ 
$GUA$ & \multicolumn{1}{c}{$(1,0,0,0,0,0)$} & \multicolumn{1}{c}{$3.155$} & 
\multicolumn{1}{c}{$Inconclusive$} &  &  &  &  &  &  & \multicolumn{1}{c}{}
&  \\ 
$HAI$ & \multicolumn{1}{c}{$(2,1,1,2,0,0)$} & \multicolumn{1}{c}{$4.164$} & 
\multicolumn{1}{c}{$CI$} & $-180.42$ & $42.822$ & $-3.804$ & $2.411^{b)}$ & $%
-1.344^{a)}$ & $0.483$ & \multicolumn{1}{c}{$278.24$} & $-0.309^{a)}$ \\ 
$HON$ & \multicolumn{1}{c}{$(1,0,0,2,0,1)$} & \multicolumn{1}{c}{$6.189$} & 
\multicolumn{1}{c}{$CI$} & $91.425$ & $-9.406$ & $0.646$ & $-28.638^{a)}$ & $%
-0.058$ & $0.233$ & \multicolumn{1}{c}{$1450.92$} & $-1.472^{a)}$ \\ 
$JAM$ & \multicolumn{1}{c}{$(2,0,0,1,0,0)$} & \multicolumn{1}{c}{$6.997$} & 
\multicolumn{1}{c}{$CI$} & $-822.2^{a)}$ & $193.12^{a)}$ & $-11.35^{a)}$ & $%
1.145$ & $0.420$ & $-1.400^{a)}$ & \multicolumn{1}{c}{$4940.78$} & $%
-0.282^{a)}$ \\ 
$MEX$ & \multicolumn{1}{c}{$(1,0,0,1,1,0)$} & \multicolumn{1}{c}{$10.267$} & 
\multicolumn{1}{c}{$CI$} & $-189.4$ & $42.557^{a)}$ & $-2.323^{a)}$ & $%
-0.334^{a)}$ & $0.084$ & $0.039$ & \multicolumn{1}{c}{$9508.1$} & $%
-0.835^{a)}$ \\ 
$NIC$ & \multicolumn{1}{c}{$(4,4,4,3,3,0)$} & \multicolumn{1}{c}{$2.626$} & 
\multicolumn{1}{c}{$NOT\ CI$} &  &  &  &  &  &  & \multicolumn{1}{c}{} &  \\ 
$PAN$ & \multicolumn{1}{c}{$(1,0,0,0,0,0)$} & \multicolumn{1}{c}{$3.242$} & 
\multicolumn{1}{c}{$Inconclusive$} &  &  &  &  &  &  & \multicolumn{1}{c}{}
&  \\ 
$PAR$ & \multicolumn{1}{c}{$(1,0,0,0,0,0)$} & \multicolumn{1}{c}{$1.594$} & 
\multicolumn{1}{c}{$NOT\ CI$} &  &  &  &  &  &  & \multicolumn{1}{c}{} &  \\ 
$PERU$ & \multicolumn{1}{c}{$(1,0,0,0,0,0)$} & \multicolumn{1}{c}{$6.674$} & 
\multicolumn{1}{c}{$CI$} & $-0.033$ & $-0.633$ & $0.098$ & $-0,126$ & $%
-0.317 $ & $0.382$ & \multicolumn{1}{c}{$25.307$} & $-0.717^{a)}$ \\ 
$URU$ & \multicolumn{1}{c}{$(3,1,2,0,1,0)$} & \multicolumn{1}{c}{$2.894$} & 
\multicolumn{1}{c}{$Inconclusive$} &  &  &  &  &  &  & \multicolumn{1}{c}{}
&  \\ 
$VEN$ & \multicolumn{1}{c}{$(1,0,0,2,0,2)$} & \multicolumn{1}{c}{$6.139$} & 
\multicolumn{1}{c}{$CI$} & $264.11^{a)}$ & $-54.28^{a)}$ & $2.836^{a)}$ & $%
0.019$ & $-0.507$ & $0.356$ & \multicolumn{1}{c}{$14311.2$} & $-0.965^{a)}$
\\ \hline
\end{tabular}%
\bigskip

Narayan (2005)'s bounds test critical values: 5\% lower bound =\ 2.694; 5\%
upper bound =\ 3.829.

$^{a)}$significant parameter at 1\%; $^{b)}$ significant parameter at 5\%; $%
^{c)}$ significant parameter at 10\%; $e_{t}=\ln (CO_{2,pc});y_{t}=\ln
(GDP);y_{t}^{2}=(\ln (GDP))^{2};x_{1,t}=$ Agriculture added value to GDP$%
;~x_{2,t}=~$industry added value to GDP$;~x_{3,t}=$ services\ added value to
GDP$;EC_{t-1}:$ estimation of the cointegration error in the short run
relationship. or speed of the adjustment in the UECM for $\nabla e_{t}$

\newpage

Table 4: Results of the bounds test and estimation of the long-run
relationship for the cuadratic EKC relationship controlling for population
density and exports of primary goods $e_{t}=\beta _{0}+\beta _{1}y_{t}+\beta
_{2}y_{t}{}^{2}+\beta _{3}x_{4,t}+\beta _{4}x_{5,t}+u_{t}$

\hspace{-2cm}%
\begin{tabular}{llllccccclc}
Country & ARDL model & Bounds test & Conclusion & $\beta _{0}$ & $\beta _{1}$
& $\beta _{2}$ & $\beta _{3}$ & $\beta _{4}$ & Turning point & $EC_{t-1}$ \\ 
\hline
$ARG$ & \multicolumn{1}{c}{$(1,1,0,0,0)$} & \multicolumn{1}{c}{$2.876$} & 
\multicolumn{1}{c}{$Inconclusive$} &  &  &  &  &  & \multicolumn{1}{c}{} & 
\\ 
$BOL$ & \multicolumn{1}{c}{$(1,1,0,1,1)$} & \multicolumn{1}{c}{$4.694$} & 
\multicolumn{1}{c}{$CI$} & $-50.033$ & $11.895$ & $-0.729$ & $-0.208$ & $%
1.023$ & \multicolumn{1}{c}{$3496.7$} & $-0.599^{a)}$ \\ 
$BRA$ & \multicolumn{1}{c}{$(1,1,0,0,0)$} & \multicolumn{1}{c}{$1.400$} & 
\multicolumn{1}{c}{$NOT~CI$} &  &  &  &  &  & \multicolumn{1}{c}{} &  \\ 
$CHI$ & \multicolumn{1}{c}{$(2,2,0,0,0)$} & \multicolumn{1}{c}{$2.118$} & 
\multicolumn{1}{c}{$NOT~CI$} &  &  &  &  &  & \multicolumn{1}{c}{} &  \\ 
$COL$ & \multicolumn{1}{c}{$(1,0,0,0,0)$} & \multicolumn{1}{c}{$4.634$} & 
\multicolumn{1}{c}{$CI$} & $-54.593^{a)}$ & $12.575^{a)}$ & $-0.663^{a)}$ & $%
0.128$ & $-1.291$ & \multicolumn{1}{c}{$13218.9$} & $-0.473^{a)}$ \\ 
$COS$ & \multicolumn{1}{c}{$(1,3,0,0,0)$} & \multicolumn{1}{c}{$4.468$} & 
\multicolumn{1}{c}{$CI$} & $-91.038^{a)}$ & $20.095^{a)}$ & $-1.103^{a)}$ & $%
-0.018$ & $0.004$ & \multicolumn{1}{c}{$9038.1$} & $-0.624^{a)}$ \\ 
$CUBA^{(\ast )}$ & \multicolumn{1}{c}{$(2,0,3,0)$} & \multicolumn{1}{c}{$%
6.387$} & \multicolumn{1}{c}{$CI$} & $9.308$ & $0.468$ & $0.018$ &  & $%
-2.901^{a)}$ & \multicolumn{1}{c}{$513445.9$} & $-0.515^{a)}$ \\ 
$R.DOM$ & \multicolumn{1}{c}{$(1,2,2,3,0)$} & \multicolumn{1}{c}{$3.726$} & 
\multicolumn{1}{c}{$Inconclusive$} &  &  &  &  &  & \multicolumn{1}{c}{} & 
\\ 
$ECU$ & \multicolumn{1}{c}{$(3,0,0,0,1)$} & \multicolumn{1}{c}{$6.217$} & 
\multicolumn{1}{c}{$CI$} & $-111.81$ & $35.513^{b)}$ & $-1.960^{b)}$ & $%
-8.848$ & $-1.830$ & \multicolumn{1}{c}{$9059.4$} & $-0.238^{a)}$ \\ 
$ELSAL$ & \multicolumn{1}{c}{$(1,0,1,0,0)$} & \multicolumn{1}{c}{$4.363$} & 
\multicolumn{1}{c}{$CI$} & $-331.85$ & $80.845^{a)}$ & $-5.082^{a)}$ & $%
0.004 $ & $1.824^{a)}$ & \multicolumn{1}{c}{$2847.2$} & $0.625^{a)}$ \\ 
$GUA$ & \multicolumn{1}{c}{$(1,0,1,0,0)$} & \multicolumn{1}{c}{$2.058$} & 
\multicolumn{1}{c}{$NOT~CI$} &  &  &  &  &  & \multicolumn{1}{c}{} &  \\ 
$HAI$ & \multicolumn{1}{c}{$(1,1,1,0,0)$} & \multicolumn{1}{c}{$4.099$} & 
\multicolumn{1}{c}{$CI$} & $13.089$ & $-8.832$ & $0.708$ & $0.209$ & $%
1.495^{a)}$ & \multicolumn{1}{c}{$512.9$} & $-0.378^{a)}$ \\ 
$HON$ & \multicolumn{1}{c}{$(1,0,0,0,2)$} & \multicolumn{1}{c}{$2.626$} & 
\multicolumn{1}{c}{$NOT~CI$} &  &  &  &  &  & \multicolumn{1}{c}{} &  \\ 
$JAM$ & \multicolumn{1}{c}{$(2,2,2,0,4)$} & \multicolumn{1}{c}{$10.586$} & 
\multicolumn{1}{c}{$CI$} & $-277.69$ & $66.447$ & $-3.626$ & $0.106^{b)}$ & $%
0.294$ & \multicolumn{1}{c}{$9533.8$} & $-0.712^{a)}$ \\ 
$MEX$ & \multicolumn{1}{c}{$(1,0,0,0,2)$} & \multicolumn{1}{c}{$4.991$} & 
\multicolumn{1}{c}{$CI$} & $-125.74^{a)}$ & $27.960^{a)}$ & $-1.508^{a)}$ & $%
0.092^{b)}$ & $-0.577^{a)}$ & \multicolumn{1}{c}{$10639.2$} & $-0.502^{a)}$
\\ 
$NIC$ & \multicolumn{1}{c}{$(4,2,0,0,2)$} & \multicolumn{1}{c}{$3.167$} & 
\multicolumn{1}{c}{$Inconclusive$} &  &  &  &  &  & \multicolumn{1}{c}{} & 
\\ 
$PAN$ & \multicolumn{1}{c}{$(4,1,1,2,3)$} & \multicolumn{1}{c}{$3.010$} & 
\multicolumn{1}{c}{$Inconclusive$} &  &  &  &  &  & \multicolumn{1}{c}{} & 
\\ 
$PAR$ & \multicolumn{1}{c}{$(1,1,0,1,0)$} & \multicolumn{1}{c}{$9.040$} & 
\multicolumn{1}{c}{$CI$} & $-9.813^{a)}$ & $4.354^{a)}$ & $-0.277^{a)}$ & $%
2.499^{a)}$ & $1.043^{a)}$ & \multicolumn{1}{c}{$3851.4$} & $-0.946^{a)}$ \\ 
$PERU$ & \multicolumn{1}{c}{$(1,0,0,2,0)$} & \multicolumn{1}{c}{$10.067$} & 
\multicolumn{1}{c}{$CI$} & $-29.063$ & $3.983$ & $-0.196$ & $1.807^{a)}$ & $%
0.447^{a)}$ & \multicolumn{1}{c}{$26423.3$} & $-0.701^{a)}$ \\ 
$URU$ & \multicolumn{1}{c}{$(1,0,0,0,0)$} & \multicolumn{1}{c}{$3.422$} & 
\multicolumn{1}{c}{$Inconclusive$} &  &  &  &  &  & \multicolumn{1}{c}{} & 
\\ 
$VEN$ & \multicolumn{1}{c}{$(1,1,0,0,0)$} & \multicolumn{1}{c}{$2.628$} & 
\multicolumn{1}{c}{$Inconclusive$} &  &  &  &  &  & \multicolumn{1}{c}{} & 
\\ \hline
\end{tabular}%
\bigskip

Narayan (2005)'s bounds test critical values: 5\% lower bound =\ 2.763; 5\%
upper bound =\ 3.813.

$^{a)}$significant parameter at 1\%; $^{b)}$ significant parameter at 5\%; $%
^{c)}$ significant parameter at 10\%; $e_{t}=\ln (CO_{2,pc});y_{t}=\ln
(GDP);y_{t}^{2}=(\ln (GDP))^{2};x_{4,t}=$ exports of~primary goods;$%
~x_{5,t}= $ population density$;EC_{t-1}:$ estimation of the cointegration
error in the short run relationship. or speed of the adjustment in the UECM
for $\nabla e_{t};$ $^{(\ast )}$ The model for CUBA does not include exports
of primary goods due to the lack of data .

\newpage

Table 5: Results of the bounds test and estimation of the long-run
relationship for the cuadratic EKC relationship controlling for external
founds, external relationships and agricultural land $e_{t}=\beta _{0}+\beta
_{1}y_{t}+\beta _{2}y_{t}{}^{2}+\beta _{3}x_{6,t}+\beta _{4}x_{7,t}+\beta
_{5}x_{8,t}+\beta _{6}x_{9,t}+u_{t}$

\hspace{-4cm}%
\begin{tabular}{llllccccccclc}
Country & ARDL model & Bounds test & Conclusion & $\beta _{0}$ & $\beta _{1}$
& $\beta _{2}$ & $\beta _{3}$ & $\beta _{4}$ & $\beta _{5}$ & $\beta _{6}$ & 
Turning point & $EC_{t-1}$ \\ \hline
$ARG$ & \multicolumn{1}{c}{$(2,0,0,0,0,0,0)$} & \multicolumn{1}{c}{$6.644$}
& \multicolumn{1}{c}{$CI$} & $-60.362$ & $12.305$ & $-0.664$ & $0.010^{a)}$
& $0.013$ & $-0.068$ & $1.247^{a)}$ & \multicolumn{1}{c}{$10620.1$} & $%
-0.852^{a)}$ \\ 
$BOL$ & \multicolumn{1}{c}{$(3,3,4,0,0,4,4)$} & \multicolumn{1}{c}{$6.160$}
& \multicolumn{1}{c}{$CI$} & $-731.36$ & $186.67$ & $-12.286$ & $0.027$ & $%
-0.015$ & $-8.823$ & $14.773$ & \multicolumn{1}{c}{$1992.0$} & $0.269^{a)}$
\\ 
$BRA$ & \multicolumn{1}{c}{$(1,4,4,3,4,2,3)$} & \multicolumn{1}{c}{$11.032$}
& \multicolumn{1}{c}{$CI$} & $-480.70$ & $108.10^{c)}$ & $-5.497^{c)}$ & $%
-0.167$ & $-0.411$ & $0.515^{c)}$ & $-0.401$ & \multicolumn{1}{c}{$11820.7$}
& $0.235^{a)}$ \\ 
$CHI$ & \multicolumn{1}{c}{$(1,0,1,0,0,0,1)$} & \multicolumn{1}{c}{$2.694$}
& \multicolumn{1}{c}{$Inconclusive$} &  &  &  &  &  &  &  & 
\multicolumn{1}{c}{} &  \\ 
$COL$ & \multicolumn{1}{c}{$(1,0,0,2,0,1,1)$} & \multicolumn{1}{c}{$2.149$}
& \multicolumn{1}{c}{$NOTCI$} &  &  &  &  &  &  &  & \multicolumn{1}{c}{} & 
\\ 
$COS$ & \multicolumn{1}{c}{$(3,4,4,4,4,4,3$} & \multicolumn{1}{c}{$4.421$} & 
\multicolumn{1}{c}{$CI$} & $-47.82$ & $8.433$ & $-0.392$ & $0.005$ & $-0.585$
& $1.001$ &  & \multicolumn{1}{c}{$46929.4$} & $.0.759^{a)}$ \\ 
$CUBA^{(\ast )}\;$ & \multicolumn{1}{c}{$(2,1,0,0,3,0)$} & 
\multicolumn{1}{c}{$4.448$} & \multicolumn{1}{c}{$CI$} & $14.862$ & $-3.911$
& $0.254$ &  & $0.259$ & $0.117$ & $-0.056$ & \multicolumn{1}{c}{$2184.1$} & 
$-0.458^{a)}$ \\ 
$R.DOM$ & \multicolumn{1}{c}{$(1,1,0,0,0,0,0)$} & \multicolumn{1}{c}{$1.386$}
& \multicolumn{1}{c}{$NOT~CI$} &  &  &  &  &  &  &  & \multicolumn{1}{c}{} & 
\\ 
$ECU$ & \multicolumn{1}{c}{$(3,0,0,0,2,3,0)$} & \multicolumn{1}{c}{$7.267$}
& \multicolumn{1}{c}{$CI$} & $-572.85^{a)}$ & $135.99^{a)}$ & $-8.084^{a)}$
& $-0.046$ & $-2.306$ & $4.079$ & $-1.455$ & \multicolumn{1}{c}{$4497.6$} & $%
-0.327^{a)}$ \\ 
$ELSAL$ & \multicolumn{1}{c}{$(2,1,0,0,4,0,2)$} & \multicolumn{1}{c}{$5.684$}
& \multicolumn{1}{c}{$CI$} & $-301.72^{a)}$ & $71.29^{a)}$ & $-4.468^{a)}$ & 
$0.027^{a)}$ & $-0.225$ & $-0.910^{a)}$ & $4.984^{a)}$ & \multicolumn{1}{c}{$%
2917.6$} & $-0.407^{a)}$ \\ 
$GUA$ & \multicolumn{1}{c}{$(1,0,1,0,0,0,0)$} & \multicolumn{1}{c}{$1.338$}
& \multicolumn{1}{c}{$NOT~CI$} &  &  &  &  &  &  &  & \multicolumn{1}{c}{} & 
\\ 
$HAI$ & \multicolumn{1}{c}{$(1,1,1,1,0,0,0)$} & \multicolumn{1}{c}{$3.165$}
& \multicolumn{1}{c}{$Inconclusive$} &  &  &  &  &  &  &  & 
\multicolumn{1}{c}{} &  \\ 
$HON$ & \multicolumn{1}{c}{$(1,0,0,0,0,0,1)$} & \multicolumn{1}{c}{$4.286$}
& \multicolumn{1}{c}{$CI$} & $76.37$ & $-21.516^{c)}$ & $1.537^{b)}$ & $%
0.016 $ & $0.482^{a)}$ & $0.001$ & $-1.043^{a)}$ & \multicolumn{1}{c}{$%
1096.9 $} & $-0.784^{a)}$ \\ 
$JAM$ & \multicolumn{1}{c}{$(4,0,1,4,4,3,4)$} & \multicolumn{1}{c}{$11.032$}
& \multicolumn{1}{c}{$CI$} & $-376.69^{a)}$ & $85.37^{a)}$ & $-4.954^{a)}$ & 
$-0.028^{a)}$ & $-.0.008^{a)}$ & $-0.909^{a)}$ & $3.530^{c)}$ & 
\multicolumn{1}{c}{$5607.2$} & $-0.740^{a)}$ \\ 
$MEX$ & \multicolumn{1}{c}{$(3,4,4,3,2,2,4)$} & \multicolumn{1}{c}{$8.849$}
& \multicolumn{1}{c}{$CI$} & $-320.91^{a)}$ & $77.04^{b)}$ & $-4.231^{b)}$ & 
$0.126$ & $-0.830^{c)}$ & $0.431^{c)}$ & $-6.841^{c)}$ & \multicolumn{1}{c}{$%
8993.0$} & $-0.338^{a)}$ \\ 
$NIC$ & \multicolumn{1}{c}{$(4,4,4,2,4,4,2)$} & \multicolumn{1}{c}{$6.362$}
& \multicolumn{1}{c}{$CI$} & $21.442^{c)}$ & $-8.204^{b)}$ & $0.607^{b)}$ & $%
0.004$ & $-0.267^{a)}$ & $0.230^{a)}$ & $1.208^{a)}$ & \multicolumn{1}{c}{$%
746.2$} & $-2.979^{a)}$ \\ 
$PAN$ & \multicolumn{1}{c}{$(4,4,4,4,2,4,4)$} & \multicolumn{1}{c}{$6.125$}
& \multicolumn{1}{c}{$CI$} & $-162.68^{c)}$ & $33.39^{b)}$ & $-2.117^{c)}$ & 
$0.048$ & $0.700$ & $-0.170$ & $-6.137$ & \multicolumn{1}{c}{$?$} & $%
-1.502^{a)}$ \\ 
$PAR$ & \multicolumn{1}{c}{$(4,2,2,0,0,1,0)$} & \multicolumn{1}{c}{$2.728$}
& \multicolumn{1}{c}{$Inconclusive$} &  &  &  &  &  &  &  & 
\multicolumn{1}{c}{} &  \\ 
$PERU$ & \multicolumn{1}{c}{$(1,0,0,0,0,0,0)$} & \multicolumn{1}{c}{$4.808$}
& \multicolumn{1}{c}{$CI$} & $-34.55$ & $7.447$ & $-0.378$ & $-0.002$ & $%
-0.096$ & $-0.081$ & $-0.120$ & \multicolumn{1}{c}{$19007.7$} & $-0.597$ \\ 
$URU$ & \multicolumn{1}{c}{$(1,2,0,0,0,0,0)$} & \multicolumn{1}{c}{$3.201$}
& \multicolumn{1}{c}{$Inconclusive$} &  &  &  &  &  &  &  & 
\multicolumn{1}{c}{} &  \\ 
$VEN$ & \multicolumn{1}{c}{$(1,0,1,1,0,0,0)$} & \multicolumn{1}{c}{$4.660$}
& \multicolumn{1}{c}{$CI$} & $155.63$ & $-35.03$ & $1.87$ & $0.280^{a)}$ & $%
0.014$ & $-0.262$ & $3.400^{a)}$ & \multicolumn{1}{c}{$11687.8$} & $%
-0.685^{a)}$ \\ \hline
\end{tabular}%
\bigskip

Narayan (2005)'s bounds test critical values: 5\% lower bound =\ 2.591; 5\%
upper bound =\ 3.766.

$^{a)}$significant parameter at 1\%; $^{b)}$ significant parameter at 5\%; $%
^{c)}$ significant parameter at 10\%; $e_{t}=\ln (CO_{2,pc});y_{t}=\ln
(GDP);y_{t}^{2}=(\ln (GDP))^{2};x_{6,t}=$ Foreign Direct Investment (FDI); $%
~x_{7,t}=$ Exports$;\ x_{8,t}=\ $Imports; $x_{9,t}=\ $Agricultural land; $%
EC_{t-1}:$ estimation of the cointegration error in the short run
relationship. or speed of the adjustment in the UECM for $\nabla e_{t};\
^{(\ast )}\ $Model for CUBA does not include Foreign Direct Investment due
to the lack of data.\bigskip

\newpage

Table 6: Results of the bounds test and estimation of the long-run
relationship for the cuadratic EKC relationship controlling for renewal
energy production, population density and rural population $e_{t}=\beta
_{0}+\beta _{1}y_{t}+\beta _{2}y_{t}{}^{2}+\beta _{3}x_{10,t}+\beta
_{4}x_{5,t}+\beta _{5}x_{11,t}+u_{t}$

\hspace{-4cm}%
\begin{tabular}{llllcccccclc}
Country & ARDL model & Bounds test & Conclusion & $\beta _{0}$ & $\beta _{1}$
& $\beta _{2}$ & $\beta _{3}$ & $\beta _{4}$ & $\beta _{5}$ & Turning point
& $EC_{t-1}$ \\ \hline
$ARG$ & \multicolumn{1}{c}{$(1,0,0,0,1,0)$} & \multicolumn{1}{c}{$9.980$} & 
\multicolumn{1}{c}{$CI$} & $-43,222$ & $10.594$ & $-0.544$ & $-0.293^{a)}$ & 
$-1.122$ & $-1.098$ & \multicolumn{1}{c}{$16964.1$} & $-0.732^{a)}$ \\ 
$BOL$ & \multicolumn{1}{c}{$(1,1,0,0,0,0)$} & \multicolumn{1}{c}{$3.133$} & 
\multicolumn{1}{c}{$Inconclusive$} &  &  &  &  &  &  & \multicolumn{1}{c}{}
&  \\ 
$BRA$ & \multicolumn{1}{c}{$(1,0,1,0,0,1)$} & \multicolumn{1}{c}{$3.115$} & 
\multicolumn{1}{c}{$Inconclusive$} &  &  &  &  &  &  & \multicolumn{1}{c}{}
&  \\ 
$CHI$ & \multicolumn{1}{c}{$(1,0,0,0,0,0)$} & \multicolumn{1}{c}{$7.025$} & 
\multicolumn{1}{c}{$CI$} & $-1.552$ & $-3.422$ & $0.265^{c)}$ & $-0.220$ & $%
-1.929^{b)}$ & $2.718^{a)}$ & \multicolumn{1}{c}{$633.8$} & $-0.498^{a)}$ \\ 
$COL$ & \multicolumn{1}{c}{$(1,0,0,1,0,0)$} & \multicolumn{1}{c}{$2.800$} & 
\multicolumn{1}{c}{$Inconclusive$} &  &  &  &  &  &  & \multicolumn{1}{c}{}
&  \\ 
$COS$ & \multicolumn{1}{c}{$(1,0,0,1,0,1)$} & \multicolumn{1}{c}{$9.213$} & 
\multicolumn{1}{c}{$CI$} & $-42.189$ & $9.017$ & $-0.414$ & $-0.589^{a)}$ & $%
-0.850^{a)}$ & $0.506$ & \multicolumn{1}{c}{$54225.1$} & $-0.732^{a)}$ \\ 
$CUBA$ & \multicolumn{1}{c}{$(1,2,4,0,2,4)$} & \multicolumn{1}{c}{$16.173$}
& \multicolumn{1}{c}{$CI$} & $-68.743^{a)}$ & $-1.710$ & $0.174$ & $-0.047$
& $11.302^{a)}$ & $6.166^{a)}$ & \multicolumn{1}{c}{$138.1$} & $-1.390^{a)}$
\\ 
$R.DOM$ & \multicolumn{1}{c}{$(4,4,4,4,4,0)$} & \multicolumn{1}{c}{$12.568$}
& \multicolumn{1}{c}{$CI$} & $30.699$ & $-12.566^{c)}$ & $0.766^{c)}$ & $%
0.788^{c)}$ & $2.901^{a)}$ & $0.920$ & \multicolumn{1}{c}{$3646.4$} & $%
-0.686^{a)}$ \\ 
$ECU$ & \multicolumn{1}{c}{$(3,0,0,0,0,0)$} & \multicolumn{1}{c}{$4.566$} & 
\multicolumn{1}{c}{$CI$} & $-270.80^{b)}$ & $64.25^{b)}$ & $-3.626^{b)}$ & $%
-1.423$ & $-1.239$ & $1.667$ & \multicolumn{1}{c}{$7041.7$} & $-0.274^{a)}$
\\ 
$ELSAL$ & \multicolumn{1}{c}{$(1,0,1,0,1,0)$} & \multicolumn{1}{c}{$4.202$}
& \multicolumn{1}{c}{$CI$} & $-342.55$ & $87.777^{a)}$ & $-5.272^{a)}$ & $%
-0.061$ & $1.780^{a)}$ & $-0.010$ & \multicolumn{1}{c}{$2824.8$} & $%
-0.752^{c)}$ \\ 
$GUA$ & \multicolumn{1}{c}{$(1,0,1,0,0,0)$} & \multicolumn{1}{c}{$1.234$} & 
\multicolumn{1}{c}{$NOT~CI$} &  &  &  &  &  &  & \multicolumn{1}{c}{} &  \\ 
$HAI$ & \multicolumn{1}{c}{$(2,2,1,1,3,0)$} & \multicolumn{1}{c}{$7.066$} & 
\multicolumn{1}{c}{$CI$} & $496.01^{a)}$ & $-91.263^{a)}$ & $6.745^{a)}$ & $%
-14.681^{a)}$ & $-9.329^{a)}$ & $-12.22^{b)}$ & \multicolumn{1}{c}{$858.7$}
& $-0.437^{a)}$ \\ 
$HON$ & \multicolumn{1}{c}{$(1,0,0,2,0,0)$} & \multicolumn{1}{c}{$4.312$} & 
\multicolumn{1}{c}{$CI$} & $329.38^{b)}$ & $-87.795^{b)}$ & $6.014^{b)}$ & $%
-1.622^{a)}$ & $-0.182$ & $0.155$ & \multicolumn{1}{c}{$1479.6$} & $%
-0.436^{a)}$ \\ 
$JAM$ & \multicolumn{1}{c}{$(1,2,2,0,4,0)$} & \multicolumn{1}{c}{$8.630$} & 
\multicolumn{1}{c}{$CI$} & $-288.79^{b)}$ & $56.634^{b)}$ & $-3.269^{b)}$ & $%
-0.274^{b)}$ & $3.442^{b)}$ & $6.915^{a)}$ & \multicolumn{1}{c}{$3736.5$} & $%
-0.994^{a)}$ \\ 
$MEX$ & \multicolumn{1}{c}{$(1,0,0,0,0,0)$} & \multicolumn{1}{c}{$4.915$} & 
\multicolumn{1}{c}{$CI$} & $-146.37^{a)}$ & $33.694^{a)}$ & $-1.816^{a)}$ & $%
-0.041$ & $-1.001$ & $-1.095$ & \multicolumn{1}{c}{$10448.4$} & $-0.485^{a)}$
\\ 
$NIC$ & \multicolumn{1}{c}{$(1,2,2,0,0,0)$} & \multicolumn{1}{c}{$10.018$} & 
\multicolumn{1}{c}{$CI$} & $8.811$ & $-14.435^{a)}$ & $1.036^{a)}$ & $%
-1.229^{a)}$ & $-2.77^{a)}$ & $9.921^{a)}$ & \multicolumn{1}{c}{$1154.5$} & $%
-1.089^{a)}$ \\ 
$PAN$ & \multicolumn{1}{c}{$(3,0,0,0,1,0)$} & \multicolumn{1}{c}{$5.732$} & 
\multicolumn{1}{c}{$CI$} & $74.405^{a)}$ & $-5.933^{a)}$ & $0.378^{c)}$ & $%
0.028$ & $-5.921^{a)}$ & $-7.201^{a)}$ & \multicolumn{1}{c}{$2548.4$} & $%
-1.228^{a)}$ \\ 
$PAR$ & \multicolumn{1}{c}{$(1,0,0,0,0,0)$} & \multicolumn{1}{c}{$4.629$} & 
\multicolumn{1}{c}{$CI$} & $-141.11^{a)}$ & $50.365^{a)}$ & $-3.133^{a)}$ & $%
0.971^{a)}$ & $-5.701^{a)}$ & $-13.915^{a)}$ & \multicolumn{1}{c}{$3099.8$}
& $-0.600^{a)}$ \\ 
$PERU$ & \multicolumn{1}{c}{$(1,0,0,0,0,0)$} & \multicolumn{1}{c}{$8.489$} & 
\multicolumn{1}{c}{$CI$} & $-30.358$ & $11.046^{a)}$ & $-0.622^{c)}$ & $%
-0.368^{b)}$ & $-2.411^{b)}$ & $-2.744^{b)}$ & \multicolumn{1}{c}{$7192.5$}
& $-0.736^{a)}$ \\ 
$URU$ & \multicolumn{1}{c}{$(1,0,0,1,0,0)$} & \multicolumn{1}{c}{$5.062$} & 
\multicolumn{1}{c}{$CI$} & $5.621$ & $4.590$ & $-0.180$ & $-0.542^{a)}$ & $%
-8.960^{a)}$ & $-1.241^{a)}$ & \multicolumn{1}{c}{$349095.2$} & $-0.628^{a)}$
\\ 
$VEN$ & \multicolumn{1}{c}{$(1,1,0,0,0,0)$} & \multicolumn{1}{c}{$5.520$} & 
\multicolumn{1}{c}{$CI$} & $-61.55$ & $16.09$ & $-0,857$ & $0.013$ & $-0.966$
& $-1.208^{b)}$ & \multicolumn{1}{c}{$12002.9$} & $-0.802^{a)}$ \\ \hline
\end{tabular}%
\bigskip

Narayan (2005)'s bounds test critical values: 5\% lower bound =\ 2.67; 5\%
upper bound =\ 3.78.

$^{a)}$significant parameter at 1\%; $^{b)}$ significant parameter at 5\%; $%
^{c)}$ significant parameter at 10\%; $e_{t}=\ln (CO_{2,pc});\ y_{t}=\ln
(GDP);\ y_{t}^{2}=(\ln (GDP))^{2};\;x_{10,t}=$ Renewal energy production; $%
~x_{5,t}=$ Population density$;x_{11,t}=\ $Rural population; $EC_{t-1}:$
estimation of the cointegration error in the short run relationship. or
speed of the adjustment in the UECM for $\nabla e_{t}$\bigskip

\newpage

Table 7: Results of the bounds test and estimation of the long-run
relationship for the cuadratic EKC relationship controlling for energy
consumption $e_{t}=\beta _{0}+\beta _{1}y_{t}+\beta _{2}y_{t}{}^{2}+\beta
_{3}x_{12,t}+\beta _{4}x_{13,t}+\beta _{5}x_{14,t}+\beta _{6}x_{15,t}+u_{t}$
\bigskip

\hspace{-4cm}%
\begin{tabular}{llllccccccllc}
Country & ARDL model & Bounds test & Conclusion & $\beta _{0}$ & $\beta _{1}$
& $\beta _{2}$ & $\beta _{3}$ & $\beta _{4}$ & $\beta _{5}$ & $\beta _{6}$ & 
Turning point & $EC_{t-1}$ \\ \hline
$ARG$ & \multicolumn{1}{c}{$(2,1,1,2,1,3,3)$} & \multicolumn{1}{c}{$4.935$}
& \multicolumn{1}{c}{$CI$} & $131,32^{b)}$ & $-29.334^{a)}$ & $1.654^{a)}$ & 
$0.213$ & $-0.041$ & $-0.145^{b)}$ & $0.052^{b)}$ & \multicolumn{1}{c}{$%
7094.8$} & $-0.561^{a)}$ \\ 
$BOL$ & \multicolumn{1}{c}{$(2,2,0,2,0,0,0)$} & \multicolumn{1}{c}{$3.505$}
& \multicolumn{1}{c}{$Inconclusive$} &  &  &  &  &  &  &  & 
\multicolumn{1}{c}{} &  \\ 
$BRA$ & \multicolumn{1}{c}{$(1,2,3,0,4,0,0)$} & \multicolumn{1}{c}{$13.097$}
& \multicolumn{1}{c}{$CI$} & $143.59^{b)}$ & $-29.701^{b)}$ & $1.531^{b)}$ & 
$2.083^{a)}$ & $-0.114$ & $0.164^{c)}$ & $0.040$ & \multicolumn{1}{c}{$%
16316.0$} & $-0.452^{a)}$ \\ 
$CHI$ & \multicolumn{1}{c}{$(1,0,1,4,0,0,0)$} & \multicolumn{1}{c}{$7.879$}
& \multicolumn{1}{c}{$CI$} & $109.49^{a)}$ & $-22.116^{a)}$ & $1.101^{a)}$ & 
$0.028$ & $1.836^{a)}$ & $1.369^{a)}$ & $0.213^{a)}$ & \multicolumn{1}{c}{$%
23091.0$} & $-0.419$ \\ 
$COL$ & \multicolumn{1}{c}{$(1,0,0,0,2,0,0)$} & \multicolumn{1}{c}{$2.065$}
& \multicolumn{1}{c}{$NOT\ CI$} &  &  &  &  &  &  &  & \multicolumn{1}{c}{}
&  \\ 
$COS$ & \multicolumn{1}{c}{$(1,4,0,1,0,0,0)$} & \multicolumn{1}{c}{$6.717$}
& \multicolumn{1}{c}{$CI$} & $-48.84^{a)}$ & $11.646^{c)}$ & $-0.694^{b)}$ & 
$1.033^{a)}$ & $-0.204$ & $0.286^{a)}$ & $-0.267^{a)}$ & \multicolumn{1}{c}{$%
4382.8$} & $-0.996^{a)}$ \\ 
$CUBA$ & \multicolumn{1}{c}{$(1,3,3,4,2,4,4)$} & \multicolumn{1}{c}{$7.927$}
& \multicolumn{1}{c}{$CI$} & $-106.37$ & $25.943^{b)}$ & $-1.563^{b)}$ & $%
-0.360^{c)}$ & $0.134$ & $0.003$ & $0.187^{a)}$ & \multicolumn{1}{c}{$4027.7$%
} & $-0.955^{a)}$ \\ 
$R.DOM$ & \multicolumn{1}{c}{$(1,0,0,0,0,1,0)$} & \multicolumn{1}{c}{$3.619$}
& \multicolumn{1}{c}{$Inconclusive$} &  &  &  &  &  &  &  & 
\multicolumn{1}{c}{} &  \\ 
$ECU$ & \multicolumn{1}{c}{$(3,3,1,4,4,3,4)$} & \multicolumn{1}{c}{$6.210$}
& \multicolumn{1}{c}{$CI$} & $185.36$ & $-39.82$ & $2.162$ & $-3.637$ & $%
3.540$ & $2.556^{b)}$ & $1.860^{c)}$ & \multicolumn{1}{c}{$9974.5$} & $%
-0.732^{a)}$ \\ 
$ELSAL$ & \multicolumn{1}{c}{$(1,0,1,0,0,0,0)$} & \multicolumn{1}{c}{$7.451$}
& \multicolumn{1}{c}{$CI$} & $-181.33^{a)}$ & $46.124^{a)}$ & $-2.826^{a)}$
& $0.454^{a)}$ & $0.113^{b)}$ & $0.317^{a)}$ & $0.010$ & \multicolumn{1}{c}{$%
2648.4$} & $-1.012^{a)}$ \\ 
$GUA$ & \multicolumn{1}{c}{$(3,0,0,0,1,2,2)$} & \multicolumn{1}{c}{$4.007$}
& \multicolumn{1}{c}{$CI$} & $97.32$ & $-25.728^{c)}$ & $1.701^{c)}$ & $%
0.516^{a)}$ & $-0.316^{c)}$ & $0.553^{a)}$ & $0.030$ & \multicolumn{1}{c}{$%
1926.0$} & $-0.716^{a)}$ \\ 
$HAI$ & \multicolumn{1}{c}{$(1,1,0,0,0,4,1)$} & \multicolumn{1}{c}{$7.999$}
& \multicolumn{1}{c}{$CI$} & $-21.951$ & $5.772$ & $-0.371$ & $0.579^{a)}$ & 
$0.212^{a)}$ & $0.145$ & $-0.070$ & \multicolumn{1}{c}{$2383.2$} & $%
-0.621^{a)}$ \\ 
$HON$ & \multicolumn{1}{c}{$(1,0,0,0,0,3,0)$} & \multicolumn{1}{c}{$7.914$}
& \multicolumn{1}{c}{$CI$} & $26.904$ & $-7.140$ & $0.496$ & $0.631^{a)}$ & $%
-0.174^{a)}$ & $0.865^{a)}$ & $0.194^{a)}$ & \multicolumn{1}{c}{$1341.7$} & $%
-0.856^{a)}$ \\ 
$JAM$ & \multicolumn{1}{c}{$(3,0,0,3,0,4,3)$} & \multicolumn{1}{c}{$2.423$}
& \multicolumn{1}{c}{$NOT\ CI$} &  &  &  &  &  &  &  & \multicolumn{1}{c}{}
&  \\ 
$MEX$ & \multicolumn{1}{c}{$(1,4,2,0,1,3,0)$} & \multicolumn{1}{c}{$5.663$}
& \multicolumn{1}{c}{$CI$} & $-138.6$ & $30.021^{a)}$ & $-1.608^{a)}$ & $%
0.131$ & $-0.126^{a)}$ & $-0.037$ & $0.009$ & \multicolumn{1}{c}{$11312.9$}
& $-0.679^{a)}$ \\ 
$NIC$ & \multicolumn{1}{c}{$(1,3,0,0,0,0,4)$} & \multicolumn{1}{c}{$19.661$}
& \multicolumn{1}{c}{$CI$} & $-5.465$ & $2.233$ & $-0.187$ & $0.871^{a)}$ & $%
-0.316^{c)}$ & $0.725^{a)}$ & $0.004$ & \multicolumn{1}{c}{$390.8$} & $%
-1.146^{a)}$ \\ 
$PAN$ & \multicolumn{1}{c}{$(1,0,0,0,0,0,0)$} & \multicolumn{1}{c}{$3.455$}
& \multicolumn{1}{c}{$Inconclusive$} &  &  &  &  &  &  &  & 
\multicolumn{1}{c}{} &  \\ 
$PAR$ & \multicolumn{1}{c}{$(1,1,0,0,1,1,0)$} & \multicolumn{1}{c}{$9.758$}
& \multicolumn{1}{c}{$CI$} & $-17.98$ & $5.512$ & $-0.402$ & $0.452^{a)}$ & $%
0.290^{a)}$ & $0.300^{a)}$ & $0.061^{a)}$ & \multicolumn{1}{c}{$946.3$} & $%
-0.917^{a)}$ \\ 
$PERU$ & \multicolumn{1}{c}{$(1,0,0,0,0,0,0)$} & \multicolumn{1}{c}{$6.194$}
& \multicolumn{1}{c}{$CI$} & $-23.792$ & $4.682$ & $-0.216$ & $0.125$ & $%
-0.302$ & $0.083$ & $-0.040$ & \multicolumn{1}{c}{$50919.4$} & $-0.717^{a)}$
\\ 
$URU$ & \multicolumn{1}{c}{$(1,0,0,0,0,0,0)$} & \multicolumn{1}{c}{$1.779$}
& \multicolumn{1}{c}{$NOT~CI$} &  &  &  &  &  &  &  & \multicolumn{1}{c}{} & 
\\ 
$VEN$ & \multicolumn{1}{c}{$(3,0,0,0,0,1,0)$} & \multicolumn{1}{c}{$3.098$}
& \multicolumn{1}{c}{$Inconclusive$} &  &  &  &  &  &  &  & 
\multicolumn{1}{c}{} &  \\ \hline
\end{tabular}%
\bigskip

Narayan (2005)'s bounds test critical values: 5\% lower bound =\ 2.591; 5\%
upper bound =\ 3.766.

$^{a)}$significant parameter at 1\%; $^{b)}$ significant parameter at 5\%; $%
^{c)}$ significant parameter at 10\%; $e_{t}=\ln (CO_{2,pc});y_{t}=\ln
(GDP);y_{t}^{2}=(\ln (GDP))^{2};\ x_{12,t}=$ Diesel consumption $~x_{13,t}=$
Electricity consumption$;\ x_{14,t}=~$Gasoline consumption; $\ x_{14,t}=~$%
Fuel consumption; $EC_{t-1}:$ estimation of the cointegration error in the
short run relationship. or speed of the adjustment in the UECM for $\nabla
e_{t}$\bigskip \bigskip

\bigskip

\end{document}